\documentclass[twocolumn]{aastex701}
\usepackage{graphicx}
\usepackage{amsmath}
\usepackage{amssymb}
\usepackage{hyperref}
\usepackage{subfigure}
\usepackage{cases}
\usepackage{mathrsfs}
\usepackage{threeparttablex}
\usepackage{longtable}
\usepackage{amssymb}
\usepackage{pifont}
\usepackage{epstopdf}
\usepackage{xcolor}
\usepackage[all]{hypcap}
\usepackage{mwe,tikz}
\usepackage{bm}
\usepackage{multirow}
\usepackage{xspace}
\usepackage{rotating}
\usepackage{CJK}
\usepackage{url}
\usepackage{upgreek}
\usepackage{booktabs}
\usepackage{textcomp}
\usepackage{makecell}
\usepackage{enumitem}
\usepackage{csquotes}
\usepackage{tabularx}

\newcommand{\e}{\text{e}}

\newcommand{\p}{\partial}

\begin{document}

\title{Spectral Hardening Reveals Afterglow Emergence in Long-Duration Fast X-ray Transients: A Case Study of GRB 250404A/EP250404a}

\correspondingauthor{Bin-Bin Zhang; Yi-Han Iris Yin; Yuan Fang}
\email{bbzhang@nju.edu.cn; iris.yin@smail.nju.edu.cn; fangyuan@ynu.edu.cn}

\author[0000-0002-5596-5059]{Yi-Han Iris Yin}
\affiliation{School of Astronomy and Space Science, Nanjing University, Nanjing 210093, China}
\affiliation{Key Laboratory of Modern Astronomy and Astrophysics (Nanjing University), Ministry of Education, China}
\email{iris.yin@smail.nju.edu.cn}

\author[0009-0006-1010-1325]{Yuan Fang}
\affiliation{South-Western Institute for Astronomy Research, Yunnan University, Kunming, Yunnan 650504, China}
\email{fangyuan@ynu.edu.cn}

\author[0000-0003-4111-5958]{Bin-Bin Zhang}
\affiliation{School of Astronomy and Space Science, Nanjing University, Nanjing 210093, China}
\affiliation{Key Laboratory of Modern Astronomy and Astrophysics (Nanjing University), Ministry of Education, China}
\email{bbzhang@nju.edu.cn}

\author[0000-0002-2191-7286]{Chen Deng} 
\affiliation{School of Astronomy and Space Science, Nanjing University, Nanjing 210093, China}
\affiliation{Key Laboratory of Modern Astronomy and Astrophysics (Nanjing University), Ministry of Education, China}
\email{dengchen@smail.nju.edu.cn}

\author[0000-0002-5485-5042]{Jun Yang}
\affiliation{School of Astronomy and Space Science, Nanjing University, Nanjing 210093, China}
\affiliation{Key Laboratory of Modern Astronomy and Astrophysics (Nanjing University), Ministry of Education, China}
\email{jyang@smail.nju.edu.cn}

\author[0009-0009-2083-1999]{Run-Chao Chen} 
\affiliation{School of Astronomy and Space Science, Nanjing University, Nanjing 210093, China}
\affiliation{Key Laboratory of Modern Astronomy and Astrophysics (Nanjing University), Ministry of Education, China}
\email{chrczxx@smail.nju.edu.cn}

\author{Yuan Liu}
\affiliation{National Astronomical Observatories, Chinese Academy of Sciences, Beijing 100101, China}
\email{liuyuan@bao.ac.cn}

\author[0000-0001-8278-2955]{Yehao Cheng}
\affiliation{South-Western Institute for Astronomy Research, Yunnan University, Kunming, Yunnan 650504, China}
\email{yhcheng@mail.ynu.edu.cn}

\author{Dong Xu}
\affiliation{National Astronomical Observatories, Chinese Academy of Sciences, Beijing 100101, China}
\affiliation{Altay Astronomical Observatory, Altay, Xinjiang 836500, People’s Republic of China}
\email{dxu@nao.cas.cn}

\author[0000-0002-7334-2357]{Xiaofeng Wang}
\affiliation{Physics Department, Tsinghua University, Beijing, 100084, China}
\email{wang_xf@mail.tsinghua.edu.cn}

\author[0000-0001-5012-2362]{Rongfeng Shen}
\affiliation{School of Physics and Astronomy, Sun Yat-sen University, Zhuhai 519082, China} 
\email{shenrf3@mail.sysu.edu.cn}

\author[0000-0002-4205-0933]{Rui-Zhi Li} 
\affiliation{Yunnan Observatories, Chinese Academy of Sciences, Kunming 650216, People's Republic of China} 
\affiliation{University of Chinese Academy of Sciences, Beijing 101408, People's Republic of China} 
\affiliation{Center for Astronomical Mega-Science, Chinese Academy of Sciences, Beijing 100012, People's Republic of China}
\email{liruizhi@ynao.ac.cn}

\author[0000-0002-7077-7195]{Jirong Mao} 
\affiliation{Yunnan Observatories, Chinese Academy of Sciences, Kunming 650216, People's Republic of China} 
\affiliation{Center for Astronomical Mega-Science, Chinese Academy of Sciences, Beijing 100012, People's Republic of China} 
\affiliation{Key Laboratory for the Structure and Evolution of Celestial Objects, Chinese Academy of Sciences, Kunming 650216, People's Republic of China}
\email{jirongmao@mail.ynao.ac.cn}

\author[0000-0002-0096-3523]{Wen-Xiong Li}
\affiliation{National Astronomical Observatories, Chinese Academy of Sciences, Beijing 100101, China}
\email{liwx@bao.ac.cn}

\author{Alberto Javier Castro-Tirado}
\affiliation{Instituto de Astrof\'isica de Andaluc\'ia (IAA-CSIC), Glorieta de la Astronom\'ia s/n, E-18008, Granada, Spain}
\affiliation{Departamento de Ingenier\'ia de Sistemas y Autom\'atica, Escuela de Ingenier\'ias, Universidad de M\'alaga, C\/. Dr. Ortiz Ramos s\/n, E-29071, M\'alaga, Spain}
\email{ajct@iaa.es}

\author{Weihua Lei}
\affiliation{Department of Astronomy, School of Physics, Huazhong University of Science and Technology, Wuhan, Hubei 430074, China}
\email{leiwh@hust.edu.cn}

\author[0009-0002-7730-3985]{Shao-Yu Fu}
\affiliation{Department of Astronomy, School of Physics, Huazhong University of Science and Technology, Wuhan, Hubei 430074, China}
\email{syfu@hust.edu.cn}

\author[0000-0001-6374-8313]{Yuan-Pei Yang}
\affiliation{South-Western Institute for Astronomy Research, Yunnan University, Kunming, Yunnan 650504, China}
\affiliation{Purple Mountain Observatory, Chinese Academy of Sciences, Nanjing, 210023, China}
\email{ypyang@ynu.edu.cn}

\author[0009-0001-8155-7905]{Shuai-Qing Jiang}
\affiliation{National Astronomical Observatories, Chinese Academy of Sciences, Beijing 100101, China}
\affiliation{School of Astronomy and Space Science, University of Chinese Academy of Sciences, Beijing 100049, China}
\email{sqjiang@bao.ac.cn}

\author{Jie An}
\affiliation{National Astronomical Observatories, Chinese Academy of Sciences, Beijing 100101, China}
\affiliation{School of Astronomy and Space Science, University of Chinese Academy of Sciences, Beijing 100049, China}
\email{anjie@nao.cas.cn}

\author[0009-0004-4767-3146]{Chun Chen} 
\affiliation{School of Physics and Astronomy, Sun Yat-sen University, Zhuhai 519082, China} 
\affiliation{CSST Science Center for the Guangdong-Hong Kong-Macau Greater Bay Area, Sun Yat-sen University, Zhuhai 519082, China}
\email{chench386@mail2.sysu.edu.cn}

\author[0009-0003-5592-3734]{Zhong-Nan Dong} 
\affiliation{School of Physics and Astronomy, Sun Yat-sen University, Zhuhai 519082, China} 
\affiliation{CSST Science Center for the Guangdong-Hong Kong-Macau Greater Bay Area, Sun Yat-sen University, Zhuhai 519082, China} 
\affiliation{National Astronomical Observatories, Chinese Academy of Sciences, Beĳing 100101, China}
\email{dongzhn@mail2.sysu.edu.cn}

\author[0000-0002-8109-7152]{Guowang Du}
\affiliation{South-Western Institute for Astronomy Research, Yunnan University, Kunming, Yunnan 650504, China}
\email{dugking@ynu.edu.cn}

\author{Ali Esamdin} 
\affiliation{Xinjiang Astronomical Observatory, Chinese Academy of Sciences, Urumqi, Xinjiang, 830011, China} 
\affiliation{School of Astronomy and Space Science, University of Chinese Academy of Sciences, Beijing 100049, China}
\email{aliyi@xao.ac.cn}

\author{Zhou Fan}
\affiliation{National Astronomical Observatories, Chinese Academy of Sciences, Beijing 100101, China}
\email{zfan@nao.cas.cn}

\author[0000-0002-1530-2680]{Haicheng Feng} 
\affiliation{Yunnan Observatories, Chinese Academy of Sciences, Kunming 650216, People's Republic of China} 
\affiliation{Center for Astronomical Mega-Science, Chinese Academy of Sciences, Beijing 100012, People's Republic of China} 
\affiliation{Key Laboratory for the Structure and Evolution of Celestial Objects, Chinese Academy of Sciences, Kunming 650216, People's Republic of China}
\email{hcfeng@ynao.ac.cn}

\author{Lu Feng} 
\affiliation{National Astronomical Observatories, Chinese Academy of Sciences, Beijing 100101, China}
\email{jacobfeng@bao.ac.cn}

\author{Emilio Fern\'andez-Garc\'ia}
\affiliation{Instituto de Astrof\'isica de Andaluc\'ia (IAA-CSIC), Glorieta de la Astronom\'ia s/n, E-18008, Granada, Spain}
\email{emifdez@iaa.es}

\author{Xing Gao} 
\affiliation{Xinjiang Astronomical Observatory, Chinese Academy of Sciences, Urumqi, Xinjiang, 830011, China}
\email{gaoxing@nao.cas.cn}

\author{Maria Gritsevich}
\affiliation{Faculty of Science, University of Helsinki, Gustav Hällströmin katu 2, FI-00014, Finland}
\affiliation{Institute of Physics and Technology, Ural Federal University, Mira str. 19, 620002 Ekaterinburg, Russia}
\email{maria.gritsevich@helsinki.fi}

\author{Wei-Jian Guo} 
\affiliation{National Astronomical Observatories, Chinese Academy of Sciences, Beijing 100101, China}
\email{guowj@bao.ac.cn}

\author[0000-0002-0779-1947]{Jingwei Hu} 
\affiliation{National Astronomical Observatories, Chinese Academy of Sciences, Beijing 100101, China}
\email{hujingwei@nao.cas.cn}

\author{You-Dong Hu}
\affiliation{Faculty of Science, Guangxi University, 100 East Daxue Road, Xixiangtang, Nanning 530004, China}
\email{huyoudong0772@hotmail.com}

\author[0009-0000-2190-6600]{Yanlong Hua}
\affiliation{Purple Mountain Observatory, Chinese Academy of Sciences, Nanjing, 210023, China}
\affiliation{School of Astronomy and Space Sciences, University of Science and Technology of China, 230026, Hefei, People's Republic of China}
\email{ylhua@pmo.ac.cn}

\author[0009-0003-9229-9942]{Abdusamatjan Iskandar}
\affiliation{Xinjiang Astronomical Observatory, Chinese Academy of Sciences, Urumqi, Xinjiang, 830011, China} 
\affiliation{School of Astronomy and Space Science, University of Chinese Academy of Sciences, Beijing 100049, China}
\email{abudu@xao.ac.cn}

\author{Junjie Jin}
\affiliation{National Astronomical Observatories, Chinese Academy of Sciences, Beijing 100101, China}
\email{jjjin@bao.ac.cn}

\author[0000-0002-0656-075X]{Niu Li} 
\affiliation{National Astronomical Observatories, Chinese Academy of Sciences, Beijing 100101, China}
\email{liniu@bao.ac.cn}

\author[0000-0001-6820-1683]{Xia Li} 
\affiliation{School of Physics and Astronomy, Sun Yat-sen University, Zhuhai 519082, China} 
\affiliation{CSST Science Center for the Guangdong-Hong Kong-Macau Greater Bay Area, Sun Yat-sen University, Zhuhai 519082, China}
\email{lixia76@mail2.sysu.edu.cn}

\author[0000-0001-7140-1950]{Ziwei Li}
\affiliation{South-Western Institute for Astronomy Research, Yunnan University, Kunming, Yunnan 650504, China}
\email{lzw@ynu.edu.cn}

\author[0000-0002-3134-9526]{Jia-Qi Lin} 
\affiliation{School of Physics and Astronomy, Sun Yat-sen University, Zhuhai 519082, China} 
\affiliation{CSST Science Center for the Guangdong-Hong Kong-Macau Greater Bay Area, Sun Yat-sen University, Zhuhai 519082, China}
\email{linjq63@mail2.sysu.edu.cn}

\author[0000-0002-0409-5719]{Dezi Liu}
\affiliation{South-Western Institute for Astronomy Research, Yunnan University, Kunming, Yunnan 650504, China}
\email{adzliu@ynu.edu.cn}

\author[0000-0002-7420-6744]{Jinzhong Liu}
\affiliation{Xinjiang Astronomical Observatory, Chinese Academy of Sciences, Urumqi, Xinjiang, 830011, China} 
\affiliation{School of Astronomy and Space Science, University of Chinese Academy of Sciences, Beijing 100049, China}
\email{liujinzh@xao.ac.cn}

\author[0009-0007-3491-7086]{Qichun Liu}
\affiliation{Physics Department, Tsinghua University, Beijing, 100084, China}
\email{lqc22@mails.tsinghua.edu.cn}

\author[0000-0003-1295-2909]{Xiaowei Liu}
\affiliation{South-Western Institute for Astronomy Research, Yunnan University, Kunming, Yunnan 650504, China}
\email{x.liu@ynu.edu.cn}

\author{Xing Liu}
\affiliation{National Astronomical Observatories, Chinese Academy of Sciences, Beijing 100101, China}
\affiliation{School of Astronomy and Space Science, University of Chinese Academy of Sciences, Beijing 100049, China}
\email{liuxing@nao.cas.cn}

\author[0000-0002-7517-326X]{Daniele B. Malesani}
\affiliation{Cosmic Dawn Center (DAWN), Denmark}
\affiliation{Niels Bohr Institute, University of Copenhagen, Jagtvej 128, 2200 Copenhagen N, Denmark}
\affiliation{Department of Astrophysics/IMAPP, Radboud University, 6525 AJ Nijmegen, The Netherlands}
\email{daniele.malesani@nbi.ku.dk}

\author{Ignacio P\'erez-Garc\'ia}
\affiliation{Instituto de Astrof\'isica de Andaluc\'ia (IAA-CSIC), Glorieta de la Astronom\'ia s/n, E-18008, Granada, Spain}
\email{ipg@iaa.es}
 
\author[0000-0002-9615-1481]{Hui Sun}
\affiliation{National Astronomical Observatories, Chinese Academy of Sciences, Beijing 100101, China}
\email{hsun@nao.cas.cn}

\author{Xue-Feng Wu}
\affiliation{Purple Mountain Observatory, Chinese Academy of Sciences, Nanjing, 210023, China}
\email{xfwu@pmo.ac.cn}

\author[0009-0004-2243-8289]{Yun-Ao Xiao} 
\affiliation{National Astronomical Observatories, Chinese Academy of Sciences, Beijing 100101, China}
\email{xiaoya@nao.cas.cn}

\author{Ding-Rong Xiong}
\affiliation{Yunnan Observatories, Chinese Academy of Sciences, Kunming 650216, People's Republic of China}
\email{xiongdingrong@ynao.ac.cn}

\author{Shengyu Yan} 
\affiliation{Physics Department, Tsinghua University, Beijing, 100084, China}
\email{yansy19@mails.tsinghua.edu.cn}

\author[0009-0002-3014-0277]{Beibei Zhang} 
\affiliation{School of Physics and Astronomy, Beijing Normal University, Beijing 100875, People's Republic of China}
\email{bbz@mail.bnu.edu.cn}

\author[0000-0002-2510-6931]{Jinghua Zhang}
\affiliation{South-Western Institute for Astronomy Research, Yunnan University, Kunming, Yunnan 650504, China}
\email{zhang_jh@ynu.edu.cn}

\author{Haichang Zhu} 
\affiliation{Physics Department, Tsinghua University, Beijing, 100084, China}
\email{tomcthulhu@gmail.com}

\author{Zipei Zhu}
\affiliation{National Astronomical Observatories, Chinese Academy of Sciences, Beijing 100101, China}
\email{zpzhu@nao.cas.cn}

\author{Hu Zou} 
\affiliation{National Astronomical Observatories, Chinese Academy of Sciences, Beijing 100101, China}
\email{zouhu@bao.ac.cn}

\author{Weimin Yuan}
\affiliation{National Astronomical Observatories, Chinese Academy of Sciences, Beijing 100101, China}
\email{wmy@nao.cas.cn}

\author[0000-0002-9725-2524]{Bing Zhang}
\affiliation{Department of Physics, University of Hong Kong, Pokfulam Road, Hong Kong, China}
\affiliation{Nevada Center for Astrophysics and Department of Physics and Astronomy, University of Nevada Las Vegas, NV 89154, USA}
\email{bzhang1@hku.hk}

\begin{abstract}

The prompt emission and afterglow phases of gamma-ray bursts (GRBs) have been extensively studied, yet the transition between these two phases remains inadequately characterized due to limited multiwavelength observational coverage. Among the recent growing samples of fast X-ray transients observed by Einstein Probe (EP), a subgroup of GRBs are captured with long-duration X-ray emission, potentially containing featured evolution from prompt emission to the afterglow phase. In this Letter, we present a detailed analysis of GRB 250404A/EP250404a, a bright fast X-ray transient detected simultaneously by EP and the Fermi Gamma-ray Burst Monitor in X-rays and gamma rays. Its continuous X-ray emission reveals a long-duration tail, accompanied by distinct spectral evolution manifested by the spectral index $\alpha_{\rm X}$ with an initial softening, followed by an evident hardening, eventually reaching a plateau at the value of $\sim$ -2. Early optical and near-infrared observations enable broadband modeling with forward- and reverse-shock components, confirming that the X-ray hardening signals the emergence of the external-shock afterglow. From this spectral hardening we infer that the prompt phase in soft X-rays lasted $\sim300\;\mathrm{s}$, which is more than 3 times longer than the gamma-ray $T_{90}$. This well-tracked soft-hard-flat spectral pattern provides a clear indication of afterglow emergence from the fading prompt emission and offers a practical criterion for identifying a distinct population of GRBs among fast X-ray transients, even when the detection of the gamma-ray counterpart or obvious temporal break is absent. 

\end{abstract}

\section{Introduction} \label{sec:intro}

The afterglow phase of a gamma-ray burst (GRB) is conventionally defined as the stage that follows the end of the prompt emission \citep{2018pgrb.book.....Z}. From a theoretical perspective, the prompt emission is attributed to internal energy dissipation within the relativistic jet \citep{1994ApJ...430L..93R, 1994ApJ...427..708P, 1997MNRAS.287..110S, 2006ApJ...642..354Z}, while the afterglow arises from the interaction between the ejecta and the circumburst medium \citep{1993ApJ...418L...5P, 1997ApJ...476..232M}. Observationally, the prompt emission typically exhibits rapid emission with erratic variability in the sub-MeV range, which is detected by GRB-triggering instruments \citep{2014IJMPD..2330002Z}. In contrast, the afterglow is distinguished by broadband spectra following a broken power-law shape and light curves displaying multisegment broken power-law behavior, usually captured through multiwavelength follow-up observations \citep{1998ApJ...497L..17S}. Extensive studies have explored the rich observational features of prompt emission, including its duration \citep{1993ApJ...413L.101K}, spectral components \citep{2011ApJ...730..141Z}, and spectral evolution \citep{2006ApJ...637..869M, 2021A&A...656A.134G}. Meanwhile, significant efforts have been devoted to understanding the afterglow phase, which is generally well explained by synchrotron radiation originating from the external forward shock (FS) or the external reverse shock (RS) during the RS crossing stage \citep{1998ApJ...497L..17S, 2005ApJ...628..315Z, 2003ApJ...595..950Z}.

However, the transition from the prompt emission phase to the afterglow phase is often not fully traced owing to a combination of instrumental, observational, and physical challenges. Instrumentally, the narrow field of view of most telescopes, except for GRB-triggering detectors, limits opportunities for obtaining simultaneous multiwavelength coverage during the prompt emission. Follow-up observations intended to capture the afterglow often begin with delays, resulting in incomplete coverage of the early emission period. Observationally, the curvature effect at the end of the prompt phase causes the emission to progressively soften and fade over time \citep{1996ApJ...473..998F, 2000ApJ...541L..51K, 2004ApJ...614..284D}, shifting it outside the energy range or below the sensitivity threshold of gamma-ray detectors. Physically, in the initial hours after the burst, the observed emission may be a complex superposition of multiple components \citep{2006ApJ...642..354Z}, including internal dissipation from late central engine activity and emission from the external RS and the external FS. As a result, it remains challenging to pinpoint when the afterglow emission from external shocks starts to appear and eventually dominates following the decay of the prompt emission.

The Einstein Probe (EP) mission \citep{Yuan2025}, with its wide field of view, has detected dozens of fast X-ray transients, several of which exhibit prompt emission counterparts in gamma rays \citep{2024ApJ...975L..27Y, 2025NatAs...9..564L, 2025arXiv250304306J}, confirming their origin from the GRB internal energy dissipation of the relativistic jet. Notably, most of the EP-detected GRBs display soft X-ray tails that extend well beyond the duration of the gamma-ray emission, offering a valuable opportunity to study the spectral evolution during the late prompt emission and the transition to the afterglow emission. Crucially, the onboard trigger of the EP Wide-field X-ray Telescope (WXT) enables rapid localization with an accuracy of a few arcminutes, which is further refined to several arcseconds by the automatic follow-up observation from the EP Follow-up X-ray Telescope (FXT) within minutes. This capability facilitates timely multiwavelength follow-up observations by ground-based observatories during the early afterglow phase, concurrent with ongoing X-ray detection. Such synergy yields critical broadband insights into the distinct emission components. As a representative example, in this Letter, we present the EP-triggered detection of GRB 250404A/EP250404a featuring rich observations covering X-ray and gamma-ray prompt emission, as well as optical and near-infrared follow-ups within a few hours. A fast flux rise is observed in both X-ray and optical bands, following the spectral hardening in the X-ray emission. We performed the afterglow fitting on the multiwavelength data with a combined FS and RS model and identified the very early afterglow emergence marked by clear spectral evolution in the X-ray band. 

This Letter is organized as follows. Section \ref{sec:obs} describes the observations and data reductions. Section \ref{sec:trans} presents the analyses and results of the prompt emission and multiwavelength afterglow modeling. In Section \ref{sec:sum}, we summarize our findings and discuss different scenarios of the transition from the prompt emission to afterglow in the long-duration fast X-ray transients.

\begin{table}
\centering
\caption{Summary of the observed properties of GRB250404A/EP250404a. All errors represent the 1$\sigma$ uncertainties.}
\label{tab:summary}
\begin{tabular}{ll}
\hline
\hline
Observed Properties & EP250404a \\
\hline
Redshift & 1.88 \\
Galactic $N_{\rm H}$ ($\rm cm^{-2}$) & $6.00\times10^{20}$ \\
\multirow{2}{*}{Intrinsic $N_{\rm H}$ ($\rm cm^{-2}$)} & $3.73_{-0.45}^{+0.35}\times10^{22}$ \\
& $6.39_{-1.31}^{+1.61}\times10^{21}$ \\
\hline
Gamma Rays (10--1000 keV) & \\
\hline
$T_{\rm 90}$ ($\rm s$) & $90.43_{-0.37}^{+0.64}$ \\
Spectral index $\alpha_{\gamma}$ & ${-1.15}_{-0.02}^{+0.02}$ \\
Peak energy (keV) & ${55.34}_{-0.51}^{+0.64}$ \\
Peak flux ($\rm erg\,cm^{-2}\,s^{-1}$) & $8.62_{-0.28}^{+0.33}\times10^{-7}$ \\
Total fluence ($\rm erg\,cm^{-2}$) & $3.37_{-0.01}^{+0.01}\times10^{-5}$ \\
Peak luminosity ($\rm erg\,s^{-1}$) & $7.81_{-0.25}^{+0.30}\times10^{51}$ \\
Isotropic energy ($\rm erg$) & $3.05_{-0.01}^{+0.01}\times10^{53}$ \\
\hline
X-Rays (0.5--10.0 keV) & \\
\hline
Duration ($\rm s$) & $\sim 300$ \\
Spectral index$^*$ $\alpha_{\rm X}$ & ${-2.59}_{-0.06}^{+0.05}$ \\
Flux$^*$ ($\rm erg\,cm^{-2}\,s^{-1}$) & $9.61_{-0.23}^{+0.26}\times10^{-9}$ \\
Total fluence$^*$ ($\rm erg\,cm^{-2}$) & $1.20_{-0.01}^{+0.01}\times10^{-6}$ \\
Isotropic energy$^*$ ($\rm erg$) & $1.09_{-0.01}^{+0.01}\times10^{52}$ \\
\hline
\hline
\end{tabular}
\begin{tablenotes}
\footnotesize
\item * The parameters are derived in the time range 130--255 s.
\end{tablenotes}
\end{table}

\begin{figure*}
 \centering
 \includegraphics[width = 0.515\textwidth]{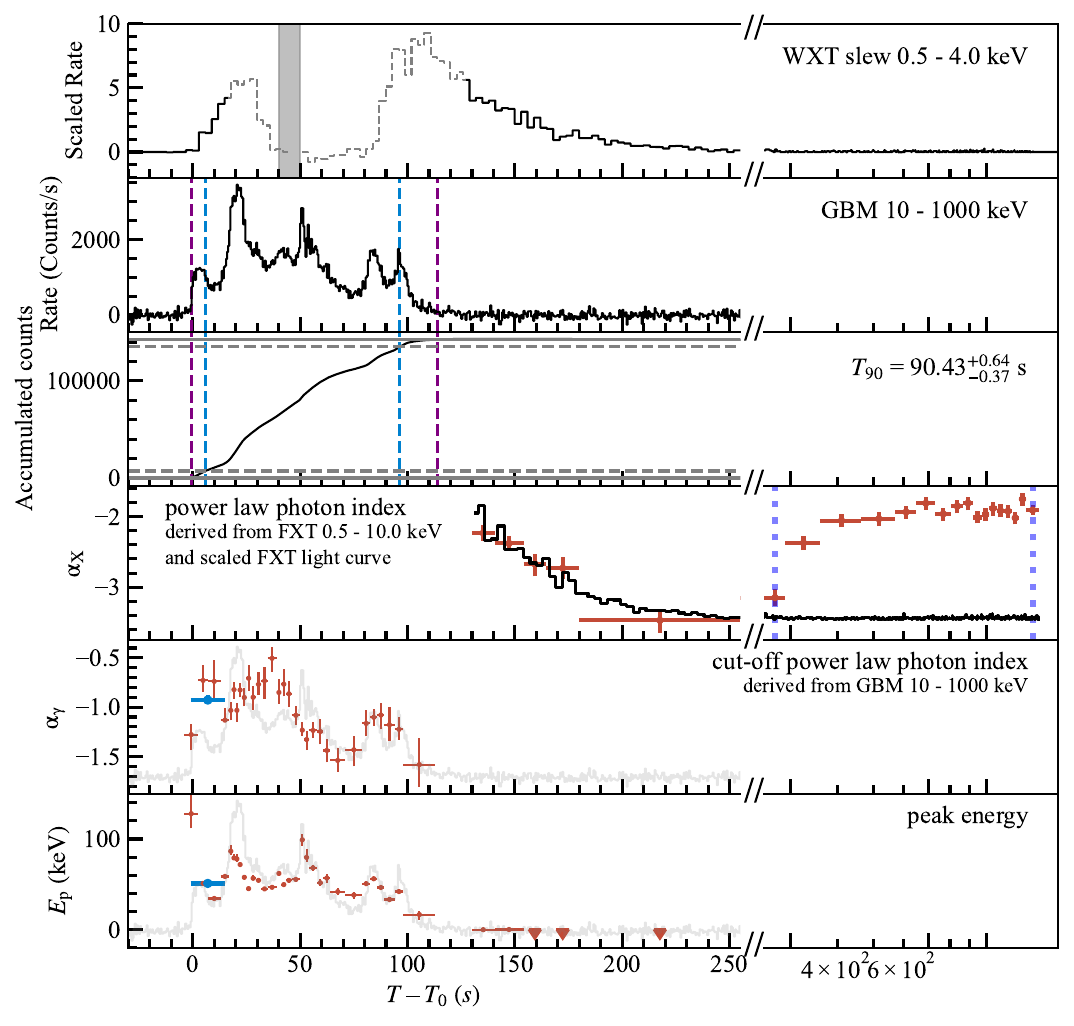}
  \includegraphics[width = 0.385\textwidth]{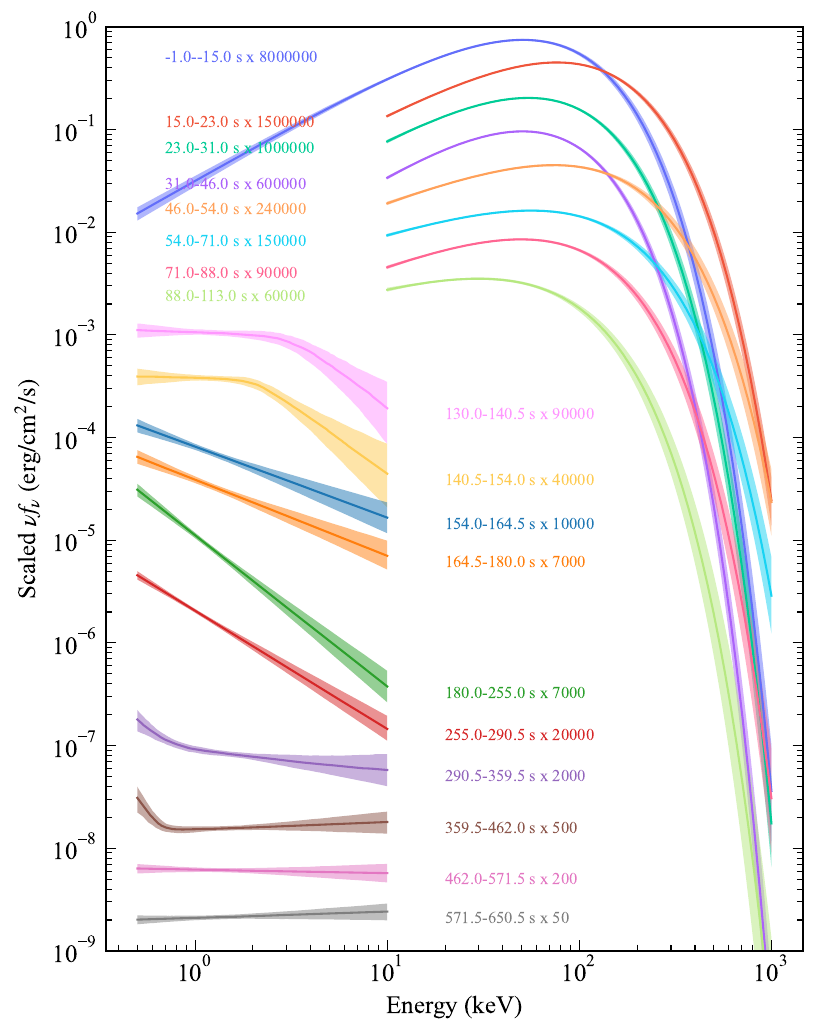}
 \caption{\textit{Left}: the observed light curve of EP250404a detected by EP/WXT in the energy range of 0.5--4.0 keV, the observed light curve of GRB 250404A detected by Fermi/GBM in the energy range of 10--1000 keV and the accumulated counts, and the spectral evolution based on the best-fit parameters of CPL, PL and SBPL models. The gray block marks the time interval where the source was outside the field of view of the detector. The dashed curve in the WXT light curve indicates a significant slew of the WXT telescope during the main emission phase, during which flux measurements may be affected. The blue dashed vertical lines represent the $T_{90}$ interval and the purple dashed vertical lines in the GBM light curve denote $T_{100}$, with solid (dashed) gray horizontal lines indicating the 0\% (5\%) and 100\% (95\%) levels, respectively. The dark blue dashed vertical lines in the FXT (fourth) panel correspond to the time interval where the spectra start to harden and reach a plateau. \textit{Right}: the evolution of the SEDs. The SEDs are derived from the spectral fittings at different time intervals listed in Tables \ref{tab:spec_fit1} and \ref{tab:spec_fit2}. All error bars mark the 1$\sigma$ confidence level.}
 \label{fig:prompt}
\end{figure*}

\section{Observations and Data Reduction}
\label{sec:obs}
\subsection{EP Observations}

EP/WXT \citep{2022hxga.book...86Y} detected the bright fast X-ray transient EP250404a on 2025 April 4 at 14:19:46 UT (referred to as $T_0$), which triggered the automatic EP/FXT \citep{2020SPIE11444E..5BC} follow-up observation at $T_0$ + 15 s \citep{2025GCN.40051....1H}. EP/FXT observed the source from $T_0$ + 130 s to 1384 s, during which the source was more precisely located at R.A. = $125.^{\circ}0601$ and decl. = $-35.^{\circ}5284$ (J2000) with an uncertainty of 10$\arcsec$ (radius, 90$\%$ cofidence, statistical and systematic) \citep{2025GCN.40085....1Y} (Figure \ref{fig:prompt}). Additional FXT follow-up detections were conducted at $T_0$ + 4324 s, $T_0$ + 44412 s and $T_0$ + 114594 s with total exposures of 2819 s, 3046 s, and 7732 s, respectively. The automatic EP/FXT follow-up observation was configured in partial window mode (FXT-A) and full-frame mode (FXT-B), and the other follow-up observations were all configured in full-frame mode.

The WXT cleaned event ﬁles and response files were generated following the standard data reduction pipelines implemented in the WXT Data Analysis Software (WXTDAS v2.10; Y. Liu et al., in preparation) and the calibration database \citep[CALDB, v1.0;][]{2025arXiv250518939C}. Source photons were extracted from a circular region with a radius of $9’$, while background photons were extracted from an annular region with inner and outer radii of $18’$ and $36’$, respectively.

FXT data were processed using the FXT Data Analysis Software (FXTDAS v1.10) developed by the EP Science Center, utilizing the latest FXT calibration database (CALDB v1.10). Given the pileup effect affecting FXT-B in full-frame mode for $\sim$ 7000 s, only the data from FXT-A in the first two EP/FXT follow-up observations are used. We estimate the pileup effect on FXT-A following the FXTDAS user guide and remove 70$\arcsec$ circle region centered on the source from $T_0$ to $T_0$ + 255 s. Within this period, the photons of the source and the background were extracted from two annular regions of the same size, with inner and outer radii of 70$\arcsec$ and 150$\arcsec$, centered on the source and a nearby clear region. For later data not affected by pileup, circular extraction regions with radii of 40$\arcsec$ and 100$\arcsec$ were used for the source and background, respectively, again centered on the source and a nearby clean region.

We note that WXT underwent a significant slew (indicated by the dashed curve in the first panel on the left of Figure~\ref{fig:prompt}) during the main emission phase. As a result, we caution that the WXT data presented in this work should be treated as qualitative references rather than precise measurements. The WXT light curve shown in Figure~\ref{fig:prompt} is therefore intended to illustrate the temporal coverage provided by EP. Our quantitative analysis of the burst properties (see Tables \ref{tab:summary}, \ref{tab:spec_fit1} and \ref{tab:spec_fit2}) relies primarily on the FXT data. 

\subsection{Fermi/GBM Observations}

The Fermi Gamma-ray Burst Monitor \citep[GBM;][]{Meegan_2009} detected GRB 250404A \citep{2025GCN.40067....1M} at $T_0$, with a calculated location consistent with that of EP250404a \citep{2025GCN.40050....1F}. We retrieved the time-tagged event data set covering the time range of GRB 250404A using the Python package \textit{heapy}\footnote{\url{https://github.com/jyangch/heapy}}. From the 12 sodium iodide detectors on board, we selected detectors n0, n1, and n9, which had the smallest viewing angles relative to the GRB source direction. In addition, we included the brightest bismuth germanium oxide detector, b0, to extend the energy coverage. Data reduction and analysis were performed with \textit{heapy}, following the standard procedures described in \citet{2011ApJ...730..141Z} and \citet{2022Natur.612..232Y}.

\subsection{Ground-based Observations}

\begin{figure}
 \centering
 \includegraphics[width = 0.45\textwidth]{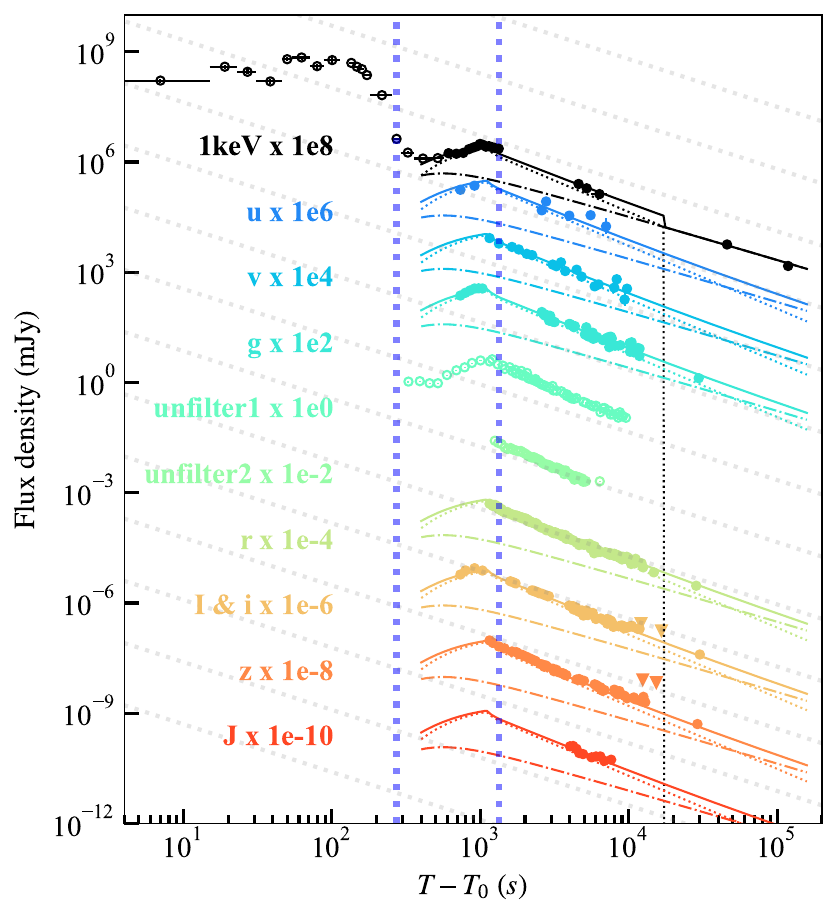}
 \caption{Multiwavelength observations of GRB 250404A/EP250404a and afterglow modeling with the best-fit parameters from the FS+RS model with $u$-, $v$-, and $g$-band correction factors. The multiwavelength data utilized in the afterglow fitting are marked with filled points, while the optical observations on clear filters that are not included in the afterglow fitting are marked with open points. The optical and near-infrared data have been corrected for Galactic extinction, which is $E(B-V)=0.073$ mag \citep{2011ApJ...737..103S}. We assume a total to selective extinction ratio of $R_V=3.1$ according to the extinction law from \citet{1999PASP..111...63F}. The inverted triangle points signify upper limits. The best-fit FS+RS, FS, and RS models are shown with solid, dashed-dotted and dotted lines, respectively. The dark blue dashed vertical lines correspond to the same time interval where the spectra start to harden and reach a plateau.}
 \label{fig:ag}
\end{figure}

We conducted ground-based follow-up observations in the optical and near-infrared bands using the Multi-channel Photometric Survey Telescope (Mephisto), the Tsinghua-Nanshan Optical Telescope (TNOT), the Sun Yat-sen University 80~cm telescope (SYSU 80cm), the Burst Observer and Optical Transient Exploring System (BOOTES)-4/María Eva Telescope (MET), and the 60/90 cm Schmidt telescope. Additional imaging data were obtained from the the Alhambra Faint Object Spectrograph and Camera (ALFOSC) mounted on the 2.56~m Nordic Optical Telescope (NOT), located at the Roque de los Muchachos Observatory, La Palma, Spain; the Half-Meter telescope (HMT) located at Nanshan Observatory, Xinjiang, China; and the JinShan 0.5~m (50D) and 1~m (100C) telescopes located at Altay Observatory, Xinjiang, China. Spectroscopic observations were carried out with the 2.4 m telescope at the Yunnan Observatories (GMG-2.4 m), covering the wavelength range of 3600--7460~\AA. A summary of these observations is provided in Figures~\ref{fig:ag} and \ref{tab:photoag} and Tables~\ref{tab:photoclear} and Figure~\ref{fig:redshift}.

\subsubsection{Photometry}

The optical counterpart of GRB 250404A/EP250404a was first detected by HMT at 14:24:59 UT on 2025 April 4, 313 s after the $T_0$. It was located at (J2000) R.A.~=~$08^{\rm hr}20^{\rm m}14.54^{\rm s}$, decl.~=~$+35^\circ 31' 41.57''$ with an uncertainty of $0.5''$ \citep{2025GCN.40052....1J}. The observation lasted approximately 2.6 hr and consisted of a series of unfiltered exposures. We calibrated the photometry of the HMT data with Gaia DR3 $G$-band reference stars.

\begin{figure}
 \centering
  \includegraphics[width = 0.45\textwidth]{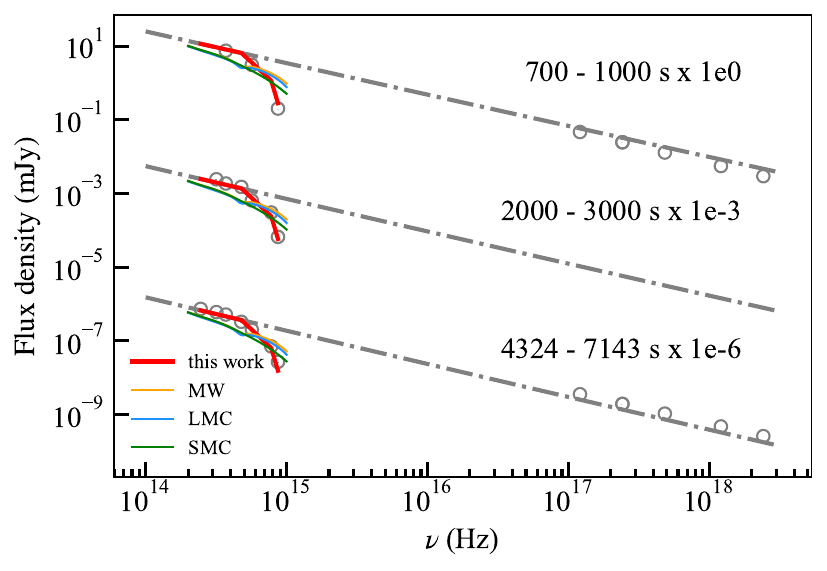}
 \caption{X-ray/optical/near-infrared afterglow specific flux density spectra of GRB 250404A/EP250404a in different time intervals. The best-fit afterglow model is indicated by dashed-dotted gray lines. The red curves correspond to the partial host galaxy extinction curve derived from the best-fit correction factors from the afterglow modeling. For comparison, the yellow, blue, and green curves show the average extinction laws of the MW, SMC, and LMC, respectively, assuming a reddening value of $E(B-V)=0.1$ mag.}
 \label{fig:fvspec}
\end{figure}

At $\sim$ 734 s after the $T_0$, Mephisto \citep{2024ApJ...969..126Y, 2024ApJ...971L...2C, 2025arXiv250315805D} started the observation in the $u$, $v$, $g$, $r$, $i$ and $z$ bands. The optical afterglow was clearly detected at R.A. = 125.$^\circ$0607, decl. = 35.$^\circ$5282 (J2000). The $u$, $g$ and $i$ bands exhibited a rapid flux rise within the first $\sim$ 300 s of the observation, followed by a decay phase characterized by differing slopes between the early and late stages. Notably, the measurements in the blue bands of Mephisto deviate from the predictions of the standard afterglow model considering the correction for Galactic extinction along the line of sight (see Figure~\ref{fig:fvspec}). This indicates the presence of additional extinction effects from the host galaxy or the surrounding medium.

We obtained a sequence of images in the $g,\,r,\,i,$ and $z$ bands using the 50D and 100C telescopes located at Altay Observatory. These observations were carried out between 0.8 and 4.6 hr after the $T_0$. Additionally, NOT/ALFOSC conducted follow-up observations in the same four filters approximately 8.3 hr after the $T_0$.

At $\sim$ $T_0$ + 1214 s, we also observed the field of the GRB 250404A/EP250404a with the 0.80m TNOT at Nanshan Station of Xinjiang Astronomical Observatory. A series of $g$-, $r$- and $i$-band images were obtained, clearly detecting the optical afterglow at R.A. = 125.$^\circ$0606, decl. = 35.$^\circ$5282 (J2000). The fluxes derived from TNOT observations are in agreement with those measured by Mephisto in the corresponding optical bands.

In the $J$ band, the Sun Yat-sen University 80cm infrared telescope started the observation at $\sim$ $T_0$ + 4021 s, obtaining 183 exposures of 20 s each. A counterpart was detected at the position of the optical afterglow in the stacked images. 

Additional observations using clear filters have also captured the long-term decay features of the afterglow. Following the EP trigger, the 0.6m BOOTES-4/MET robotic telescope at Lijiang Astronomical Observatory automatically responded to this event at $\sim$ $T_0$ + 1203 s. A series of clear-filter images were collected, revealing a source consistent with the optical afterglow position. The magnitudes were measured using Gaia DR3 $G$ band as the reference. Furthermore, the 60/90 cm Schmidt telescope at Xinglong Observatory monitored the source over an extended period from $T_0$ + 1251 s to $T_0$ + 6379 s, capturing a series of unfiltered images. The magnitudes for these observations were calibrated using Gaia DR2 $G$ band as the reference.

\subsubsection{Spectroscopy}

The spectroscopic observation was conducted using the GMG-2.4m telescope with grism \#14, which provides a wavelength coverage of 3600-7460 \AA \citep{2019RAA....19..149W,2020RAA....20..149X}. A single spectrum with an exposure time of 1800 s was obtained, beginning at $T_0$ + 3519 s. The spectrum was reduced using standard IRAF procedures \citep[v2.18.1;][]{1986SPIE..627..733T,1993ASPC...52..173T,IRAF_2_18_1}. We identified multiple metal absorption features as shown in Figure \ref{fig:redshift}, including Si IV at 1394 \AA, Si IV at 1403 \AA, Si II at 1527 \AA, C IV at 1549 \AA, Fe II at 1608 \AA, Al II at 1671 \AA, Al III at 1855 \AA, Al III at 1863 \AA, Zn II/Cr II at 2026 \AA, Zn II/Cr II at 2062 \AA, Fe II at 2344 \AA, Fe II at 2374 \AA~and Fe II at 2383 \AA. These features consistently indicate a redshift of $z$ $\sim$ 1.88.

\section{Prompt Emission to Afterglow transition}
\label{sec:trans} 
\subsection{Early-time Long-duration Light Curve}
In Figure \ref{fig:prompt}, we present the light curve of GRB 250404A/EP250404a detected by Fermi/GBM with a bin size of 0.5 s in the energy range of 10--1000 keV. The accumulated counts curve is also shown, from which we extract the $T_{90}$ interval of ${90.43}_{-0.37}^{+0.64}$ s. The prompt emission phase of GRB 250404A is indicated by the $T_{90}$ and multipulse light curve with variability in this energy range.

We also display the scaled long-duration light curve detected by EP/WXT with a bin size of 3 s in the energy range of 0.5--4.0 keV. Continuous X-ray emission was affected by the slewing motion from $T_0$ + 15 s to 130 s and was disrupted at $T_0$ + 1384 s due to the end of the observation cycle. As a result, a reliable $T_{90}$ estimation in the X-ray band for EP250404a could not be accurately derived. We also note that the second pulse of EP250404a appears broadened and extended compared to the gamma-ray emission, exhibiting a long tail with continuous emission that persists for thousands of seconds.

\subsection{Spectral Evolution}
We performed both time-integrated and time-resolved spectral fittings for Fermi/GBM and EP data using the Python package \textit{bayspec}\footnote{\url{https://github.com/jyangch/bayspec}} (an upgraded version of \textit{MySpecFit}), following the approach described in \citet{2022Natur.612..232Y} and \citet{2023ApJ...947L..11Y}. \textit{bayspec} is a Bayesian-inference-based spectral fitting tool for multidimensional and multiwavelength astrophysical data. The goodness of fit was evaluated by examining the reduced statistic STAT/dof, as described in \citet{2024ApJ...975L..27Y}. The model comparison was conducted using the Bayesian information criterion (BIC) as defined by \citet{bic_ref}. The best-fitting model parameters, along with the corresponding statistics for each time slice, are provided in Tables \ref{tab:spec_fit1} and \ref{tab:spec_fit2}. Based on the spectral fitting results, we derived the spectral evolution and the spectral energy distributions (SEDs), which are illustrated in Figure \ref{fig:prompt}. 

Before analyzing the spectral evolution in detail, we firstly divided the early high-energy emissions detected by Fermi/GBM and EP into four time intervals based on data availability and performed time-integrated spectral fittings for each episode.

\begin{figure}
 \centering
  \includegraphics[width = 0.45\textwidth]{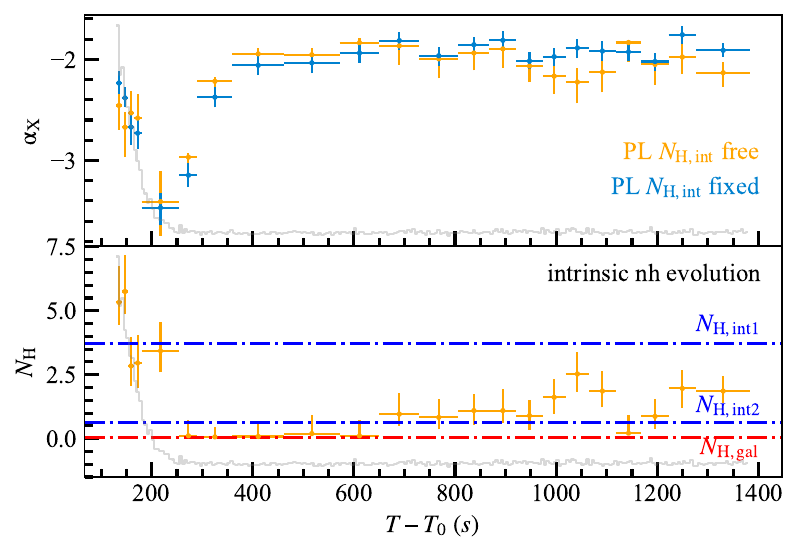}
 \caption{The evolution of spectral index $\alpha_{\rm X}$ and intrinsic absorption column density $N_{\rm H, int}$. The blue (yellow) points in the upper panel represent the best-fit spectral indices in the time-resolved spectral fittings with fixed (free) $N_{\rm H, int}$. The yellow points in the lower panels correspond to the best-fit $N_{\rm H, int}$. The two blue dotted-dashed horizontal lines mark the best-fit intrinsic absorption column density $N_{\rm H, int1}$ and $N_{\rm H, int2}$ obtained from time-integrated spectral fittings, respectively. The red dotted-dashed horizontal line denotes the Galactic absorption column density $N_{\rm H, gal}$. All error bars on data points represent their 1$\sigma$ confidence level.
}
 \label{fig:test}
\end{figure}

\begin{itemize}

 \item[(a)] \textit{EP/WXT and Fermi/GBM joint fit in $T_0$ + [-1, 15] s.} Before the time interval when EP was slewing, there was a single overlapping time range of gamma-ray and X-ray emission, spanning $T_0$ + [-1, 15] s. We performed a joint fit using an absorbed cutoff power-law (CPL) model, \textit{tbabs*ztbabs*cpl}. In this model, \textit{tbabs} and \textit{ztbabs} represent the Tuebingen-Boulder interstellar medium (ISM) absorption model \citep{Wilms2000ApJ}, with parameters for the absorption column density, $N_{\rm H}$, and redshift. We adopted a Galactic absorption column density of $N_{\rm H, gal} \sim 6.00\times10^{20}~\rm{cm^{-2}}$ and an intrinsic absorption column density of $N_{\rm H, int1}\sim 3.73\times10^{22}~\rm{cm^{-2}}$, with the latter fixed to the value obtained from the time-integrated spectral fitting in $T_0$ + [130, 255] s using the EP/FXT spectrum (see below). The joint spectrum is well fitted, yielding a spectral index of ${-0.93}_{-0.04}^{+0.05}$ and a peak energy of $50.93_{-1.62}^{+1.55}$ keV. This time interval covers the first pulse of the prompt emission detected in gamma rays, and the joint best-fit parameters represent an averaged view of the time-resolved spectral evolution during the first pulse (see below).

 \item[(b)] \textit{Fermi/GBM independent fit in $T_0$ + [-4, 113] s.} The prompt emission of GRB 250404A detected by Fermi/GBM expands from $\sim$ $T_0$ -4 s to +113 s. The time-integrated spectrum for this interval is well described by the CPL model, yielding a spectral index of ${-1.15}_{-0.02}^{+0.02}$ and a peak energy of ${55.34}_{-0.51}^{+0.64}$ keV. The isotropic energy derived from this episode is $3.05_{-0.01}^{+0.01}\times10^{53}$ erg, placing GRB 250404A within the parameter space of type II GRBs on the Amati relation \citep{2002A&A...390...81A}.

 \item[(c)] \textit{EP/FXT independent fit in $T_0$ + [130, 255] s and [255, 1384] s.} The long-duration tail of EP250404a was detected by EP/FXT from $T_0$ + 130 s to + 1384 s. Considering the possibility of time-dependent absorption in the X-ray spectra, which is not uncommon in bright GRBs \citep{2002MNRAS.330..383L, 2003ApJ...585..775P, 2003MNRAS.340..694L, 2007ApJ...654L..17C, 2021A&A...649A.135C}, we adopted different intrinsic absorption column densities for the time-resolved spectral fittings. However, the values of $N_{\rm H, int}$ for the time-resolved spectra are difficult to constrain accurately, with significant uncertainties. Thus, we determined the $N_{\rm H, int}$ from time-integrated spectra and fixed the value for time-resolved spectral fittings.

We firstly performed a time-resolved spectral fitting test using the model \textit{tbabs*ztbabs*pl}. In this model, the Galactic absorption column density $N_{\rm H, gal}$ was fixed at $6.00\times10^{20}~\rm{cm^{-2}}$, while the intrinsic absorption column density $N_{\rm H, int}$ was treated as a free parameter. We divided the time-resolved spectra, ensuring each EP/FXT spectrum contained at least 300 total accumulating photon counts. Our analysis revealed significant evolution in the best-fit values of $N_{\rm H, int}$ between the fast-decaying phase ($\sim T_0$ + [130, 255] s) and the slow-decaying phase ($\sim T_0$ + [255, 1384] s) of the X-ray emission tail. Therefore, we adopted two distinct intrinsic absorption column densities determined from independent fits in $T_0$ + [130, 255] s and [255, 1384] s: $N_{\rm H, int1}$ $\sim$ $3.73\times10^{22}$ $\rm cm^{-2}$ for the period before $T_0$ + 255 s and $N_{\rm H, int2}$ $\sim$ $6.39\times10^{21}$ $\rm cm^{-2}$ for the period afterward.

Accounting for the effect of an evolving $N_{\rm H, int}$ is nontrivial because the absorption in X-ray spectra is degenerate with the intrinsic spectral index. As shown in Figure~\ref{fig:test}, our treatment of the intrinsic absorption column density does not significantly affect the values of the spectral index, nor does it alter the overall trend of spectral index evolution.

\end{itemize}

Building on the time-integrated spectral fitting results, we performed time-resolved spectral fittings to derive the spectral evolution and SEDs, as shown in Figure~\ref{fig:prompt}. We briefly outline the time-resolved spectral fittings and analyses as follows.

\begin{enumerate}

\item \textbf{\textit{The prompt emission in gamma rays.}}

We extracted the time-resolved spectra of Fermi/GBM in the time range of $T_0$ + [-4, 113] s to explore the spectral evolution of the prompt emission in gamma rays. The time-resolved spectra were divided into two sets based on different criteria: one with a minimum of 30 average accumulated photon counts per channel and the other with a minimum of 120. The extracted spectra were fitted using the CPL model, and the spectral evolution was analyzed through the best-fit parameter. The resulting spectral evolution and SED plots are shown in Figure~\ref{fig:prompt}, where we observe that the peak energy, $E_{\rm p}$, exhibits a hard-to-soft transition during the first pulse, followed by an intensity tracking pattern \citep{1983Natur.306..451G}, with an overall declining trend. Additionally, the plots reveal that the spectral index, $\alpha_{\gamma}$, evolves with an overall softening trend, ranging from -0.51 to -1.58.

\item \textbf{\textit{The long-duration tail in X-rays.}}

The spectral evolution of the long-duration tail in X-rays was analyzed via time-resolved spectral fittings with fixed intrinsic absorption column density $N_{\rm H, int1}$ and $N_{\rm H, int2}$ determined from time-integrated spectral fittings. We employed two spectral models, an absorbed power law (PL) \textit{tbabs*ztbabs*pl} and an absorbed smooth broken power law (SBPL) \textit{tbabs*ztbabs*sbpl}, for the time-resolved spectral fittings of EP/FXT spectra. Most spectra were better described by \textit{tbabs*ztbabs*pl}. However, for the time-resolved spectra in the intervals $T_0$ + [130, 154] s and $T_0$ + [290.5, 462.0] s, both models provided comparable fits ($\Delta \rm{BIC} \textless 5$). We identified the spectral evolution of the peak energy passing through the EP/FXT 0.5--10.0 keV band within $T_0$ + [130, 154] s, as revealed by the SBPL model, which aligns with the softening trend observed in Fermi/GBM spectral fittings in $T_0$ + [-4, 113] s. Furthermore, the spectral indices $\alpha_{\rm X}$ demonstrate significant hardening after $T_0$ + 255 s, eventually reaching a plateau at $\sim-2$. Notably, the first two time-resolved spectra after $T_0$ + 255 s clearly display a transition, as seen in the SBPL model in the SED plot. This suggests the emergence of a second spectral component, distinct from the softening and fading prompt emission.

\end{enumerate}

Lastly, joint fittings were performed using EP/FXT-A and EP/FXT-B spectra for the time intervals $T_0$ + [44412, 47458] s and $T_0$ + [114594, 122326] s. We incorporated a calibration constant for EP/FXT-A to account for the systematic uncertainty between the two detectors. The X-ray flux densities were then derived for the afterglow fitting.

\begin{table*}
\centering
\small
\caption{Afterglow fitting results and corresponding fitting statistics for different models. All errors represent the 1$\sigma$ confidence level.}
\label{tab:agfit}
\begin{tabular}{cccccccccc}
\hline
\hline
Model & log$E_{\rm k,iso}$& log$\Gamma_0$& $\theta_{\rm jet}$& log$n_{18}$& $p_{\rm f}$& log$\epsilon_{\rm e, f}$& log$\epsilon_{\rm B, f}$& $p_{\rm r}$ \\
& (erg) && (deg) & $\rm (cm^{-3})$ &&&&\\
\hline
FS+RS$^*$ & ${55.38}_{-0.25}^{+0.43}$ & ${2.33}_{-0.06}^{+0.16}$ & ${7.39}_{-3.43}^{+1.01}$ & ${-0.42}_{-1.08}^{+0.50}$ & ${2.59}_{-0.01}^{+0.25}$ & ${-1.27}_{-0.48}^{+0.19}$ & ${-6.29}_{-0.28}^{+1.11}$ & ${2.77}_{-0.06}^{+0.01}$ \\
FS$^*$ & ${53.35}_{-0.02}^{+0.54}$ & ${2.89}_{-0.24}^{+0.06}$ & ${9.67}_{-0.23}^{+0.27}$ & ${3.52}_{-0.15}^{+0.65}$ & ${2.81}_{-0.01}^{+0.02}$ & ${-0.28}_{-0.55}^{+0.00}$ & ${-6.32}_{-0.54}^{+0.24}$ & - \\
FS+RS & ${53.51}_{-0.01}^{+0.03}$ & ${1.58}_{-0.02}^{+0.01}$ & ${8.44}_{-0.07}^{+0.29}$ & ${3.61}_{-0.05}^{+0.11}$ & ${2.70}_{-0.00}^{+0.01}$ & ${-0.62}_{-0.04}^{+0.01}$ & ${-6.05}_{-0.06}^{+0.06}$ & ${2.10}_{-0.00}^{+0.06}$ \\
FS & ${53.71}_{-0.21}^{+0.54}$ & ${2.91}_{-0.31}^{+0.03}$ & ${9.81}_{-0.74}^{+0.09}$ & ${3.81}_{-0.87}^{+0.25}$ & ${2.82}_{-0.01}^{+0.04}$ & ${-0.51}_{-0.58}^{+0.05}$ & ${-6.81}_{-0.04}^{+0.84}$ & - \\
\hline
Model & log$\epsilon_{\rm e, r}$& log$\epsilon_{\rm B, r}$& logv & log$f_u$& log$f_v$& log$f_g$ & $\chi^2$/dof & BIC \\
\hline
FS+RS$^*$ & ${-0.35}_{-0.08}^{+0.24}$ & ${-4.32}_{-0.79}^{+0.54}$ & ${-0.87}_{-0.00}^{+0.03}$ & ${-1.09}_{-0.04}^{+0.02}$ & ${-0.58}_{-0.01}^{+0.02}$ & ${-0.20}_{-0.00}^{+0.01}$ & 370.67/336 & -192.77 \\
FS$^*$ & - & - & ${-0.64}_{-0.01}^{+0.02}$ & ${-1.13}_{-0.02}^{+0.07}$ & ${-0.55}_{-0.03}^{+0.03}$ & ${-0.22}_{-0.02}^{+0.01}$ & 345.55/339 & 77.11 \\
FS+RS & ${-0.99}_{-0.05}^{+0.01}$ & ${-4.05}_{-0.09}^{+0.03}$ & ${-0.58}_{-0.00}^{+0.01}$ & - & - & - & 278.62/314 & 98.27 \\
FS & - & - & ${-0.51}_{-0.01}^{+0.03}$ & - & - & - & 330.84/317 & 228.74 \\
\hline
\hline
\end{tabular}
\begin{tablenotes}
\small
\item * Three constants are added to the model for the host galaxy extinction corrections on $u$, $v$ and $g$ bands.
\end{tablenotes}
\end{table*}

\subsection{Confirming Afterglow Emergence via Multiwavelength Fitting}

Corresponding to the spectral hardening phase starting at $\sim$ $T_0$ + 255 s, the flux density at 1 keV initially exhibits a brief decay, followed by a sharp rise peaking at $\sim$ $T_0$ + 1000 s. This rising flux is also observed in the $u$, $g$ and $i$ bands of Mephisto early-time optical data. Subsequently, the flux density shows a steeper slope before $\sim$ $T_0$ + 5000 s and a shallower decay at later times. These behaviors indicate the presence of a RS component emerging around $T_0$ + 1000 s, superimposed on the FS emission. 

We fitted the multiwavelength data with both FS and FS+RS models using \textit{PyFRS}\footnote{\url{https://github.com/leiwh/PyFRS}} \citep{2018pgrb.book.....Z, 2013NewAR..57..141G, 2016ApJ...816...20L, 2023ApJ...948...30Z}. 
In our modeling, we adopt the top-hat jet-type and thin-shell case. The overall evolution of the FS shell, covering the coasting, deceleration, jet-break, and Newtonian phases, is calculated numerically using a set of hydrodynamical equations as given in \citet{2000ApJ...543...90H}. In the thin shell case, the RS is Newtonian during the shock crossing phase, and its dynamical behavior is expressed with the scalings as described in \citet{2000ApJ...545..807K} (see also \citet{2013NewAR..57..141G, 2018pgrb.book.....Z}).\footnote{It should be noted that the dynamics in \textit{PyFRS} are approximate descriptions. A more rigorous treatment accounting for internal energy evolution, adiabatic losses, and total energy conservation is given by \citet{2013MNRAS.433.2107N} (see also \citet{2018pgrb.book.....Z})} The synchrotron spectra of the FS and RS are calculated following the standard broken power-law spectral model, separated by the three characteristic frequencies: the synchrotron self-absorption frequency $\nu_a$, the minimum frequency $\nu_m$, and the cooling frequency $\nu_c$. In parallel with the physical modeling, we also considered the potential host galaxy extinction affecting the blue bands. Here, we adopted two approaches: (1) fitting the afterglow without the heavily affected $u$ and $v$ bands and (2) fitting the afterglow with extinction correction factors applied to the $u$, $v$, and $g$ bands.

The fitting of the multiwavelength data is implemented by the Bayesian computation python package \textit{PyMultinest} \citep{2014A&A...564A.125B} with the log-likelihood function written as

\begin{equation}
\ln\mathcal{L}=-\frac{1}{2}\sum_{i=1}^{n}\left\{\frac{(O_i-P_i)^2}{\sigma_i^2+v^2}+\ln\left[2\pi(\sigma_i^2+v^2)\right]\right\},
\end{equation}
where $O_i$, $P_i$, and $\sigma_i$ stand for the $i$th of $n$ observed magnitudes, model-predicted magnitudes, and the uncertainties of observed magnitudes, respectively. An additional variance parameter $v$ is introduced as a scatter term, which accounts for additional uncertainty in the models and data. For upper limits, a one-sided Gaussian penalty term is applied. To measure the goodness of fit between the model and the observed data, we utilize $\chi^2$ as the statistical metric. All model parameters are allowed to vary, with broad but physically reasonable priors, enabling a comprehensive search for the best-fit solution. 

The best-fitting model parameters and their corresponding statistics for the four fitting scenarios are presented in Table \ref{tab:agfit}. Among these, only the FS+RS model with extinction correction factors accurately reproduces the multiwavelength data. The corner plot of the posterior probability distributions of the parameters are presented in Figure \ref{fig:corner}. Furthermore, the derived extinction correction factors are consistent with the specific flux density spectra shown in Figure \ref{fig:fvspec} in different time slices, which formed a steeper curve compared to the average extinction laws of Milky Way (MW), Small Magellanic Cloud (SMC) and Large Magellanic Cloud (LMC), assuming a reddening value of $E(B-V)=0.1$. 

These results confirm that the second spectral component emerging after $T_0$ + 255 s corresponds to the FS+RS afterglow emission of GRB 250404A/EP250404a.

\section{Summary and Discussion}
\label{sec:sum}
In this Letter, we present a detailed analysis of the long-duration emission from the fast X-ray transient EP250404a, whose gamma-ray counterpart was simultaneously detected as GRB 250404A (see Table \ref{tab:summary}). The wide-field detector EP/WXT captured the prompt emission phase of GRB 250404A, while the rapid automatic follow-up of EP/FXT recorded a continuous emission tail lasting for thousands of seconds after the trigger. We conducted a comprehensive spectral analysis of the long-duration emission in gamma-ray and X-ray bands, revealing a complete spectral evolution from the prompt emission to the afterglow. The spectral peak energy decreased from above 100 keV to below 0.5 keV within the first $\sim$ 154 s from the trigger. The spectral index $\alpha_{\rm X}$ showed a softening trend during the decaying tail before $T_0$ + 255 s, followed by a clear hardening, eventually reaching a plateau at $\sim$ -2. This spectral hardening coincides with a sharp flux rise at 1 keV and in the optical bands, peaking at $\sim$ $T_0$ + 1000 s. The subsequent decay in flux density is characterized by distinct slopes in the early and late stages. Our multiwavelength data are best fitted by an FS+RS afterglow model with extinction correction factors in the blue bands, confirming that the emerging second component at $T_0$ + 255 s, indicated by the spectral hardening, is the afterglow emission. These findings also provide an estimation of the prompt emission phase duration in X-rays, lasting $\sim$ 300 s, after which the X-ray emission is dominated by the afterglow.

Interestingly, spectral hardening has become less uncommon in the detection of long-duration fast X-ray transients by EP thanks to its automatic rapid follow-up capability, which enables the identification of an increasing number of events exhibiting these features. This phenomenon, characterized by the evolution of peak energy $E_{\rm p}$, softening and hardening of the spectral index $\alpha_{\rm X}$, and the appearance of spectral breaks in the SEDs, may signify a universal transition from the prompt emission phase to the afterglow phase. Benefiting from simultaneous gamma-ray observations and extensive multiwavelength follow-ups in the optical and near-infrared bands, we were able to conduct an in-depth analysis of the spectral evolution for this event. In this context, GRB 250404A/EP250404a may serve as a representative case of a specific subgroup of fast X-ray transients, distinguished by a continuous decay phase exhibiting both spectral softening and subsequent hardening. The physical mechanisms explored in this study could potentially be applicable to other similar events.

In the canonical scenario, the afterglow begins as the FS is launched into the external medium. Initially, its emission is much fainter than that of the prompt emission. As the blast wave accumulates ambient mass and the bulk Lorentz factor decreases, energy dissipation becomes more significant, and the afterglow emerges observationally, typically outshining the prompt emission at later times. Based on our findings in this case, we propose two observational indicators of the transition from prompt emission to afterglow in the X-ray band. First, a temporal break separates the steep decay (attributed to the tail of the prompt emission) from the normal decay (dominated by the afterglow). Second, the observed spectral hardening, following the initial softening associated with the prompt emission and leading into a flatter, harder afterglow spectrum, may signify a shift from prompt-dominated to afterglow-dominated emission. These findings suggest that the transition between the prompt emission and afterglow phases can occur in various forms, depending on the relative strengths and durations of the two components. We briefly summarize four typical scenarios (as illustrated in Figure \ref{fig:illus}) for the transition between prompt emission and afterglow in GRBs that are observed in long-duration fast X-ray transients.

\begin{itemize}

\item \textit{Case I}: Smooth spectral transition with a temporal break. The prompt emission and afterglow components are superimposed, with the prompt emission exhibiting a harder spectral index than the afterglow at the point when the afterglow begins to dominate at later times. In this scenario, a gradual softening of the spectral index is expected, while the light curve exhibits a clear temporal break separating the steep decay of the prompt emission tail from the normal afterglow decay.

\item \textit{Case II}: Spectral hardening with smooth temporal decay. The prompt and afterglow emissions are superimposed and of comparable intensity. The presence of the afterglow broadens the fast-decaying profile of the prompt emission. In this case, the spectral index of the prompt emission softens beyond the value of the afterglow (we take $\sim$ $-2$ for the demonstration) before the afterglow becomes dominant. This results in a characteristic soft-hard-flat pattern in the spectral index evolution, accompanied by a smoothly decaying light curve. The spectral evolution in this case is particularly informative for identifying the emergence of the afterglow. The prompt emission and afterglow components can be disentangled by fitting the time-resolved spectra during the spectral hardening phase with two power-law models, allowing for the determination of their respective fluxes. While the total decay light curve is featureless, the decay slopes of the separated components align with standard GRB expectations: a steep decay for the prompt emission and a shallower, typical decay for the afterglow.

\begin{figure}
 \centering
 \includegraphics[width = 0.45\textwidth]{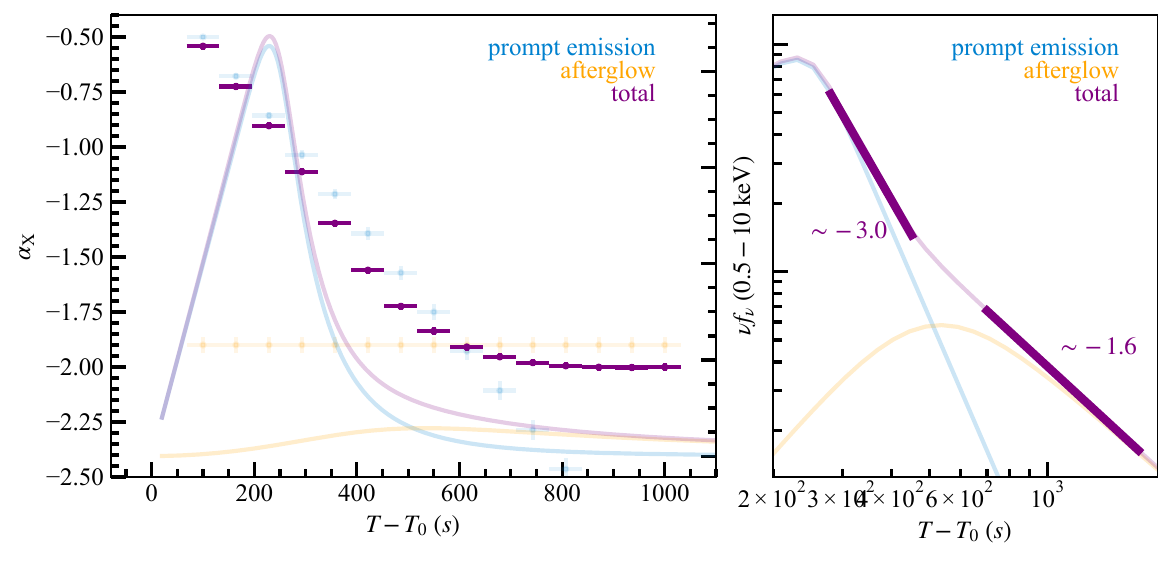}
 \includegraphics[width = 0.45\textwidth]{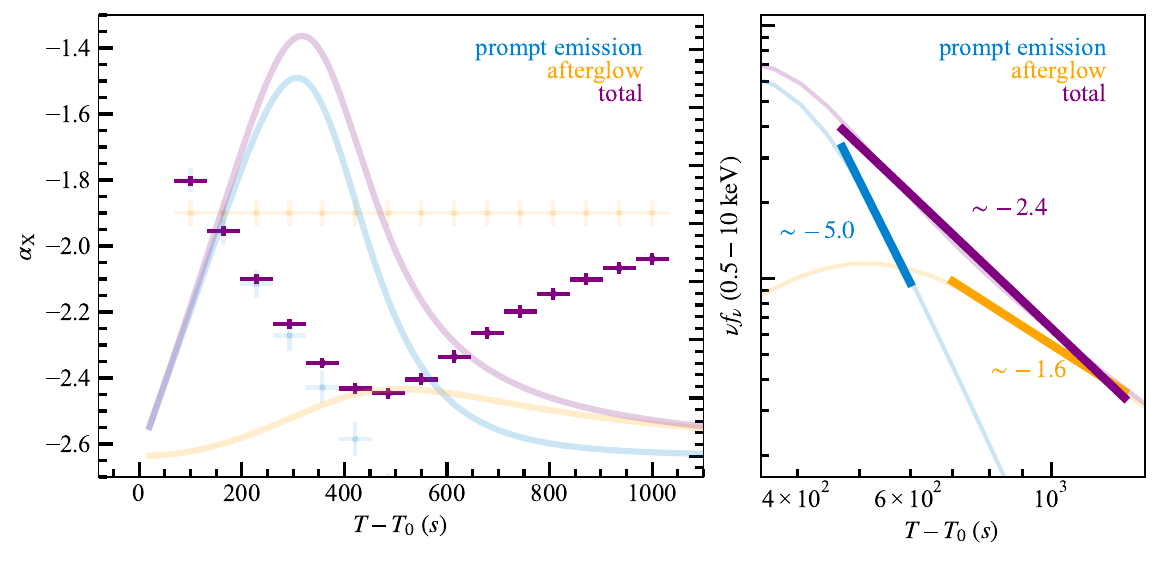}
 \includegraphics[width = 0.45\textwidth]{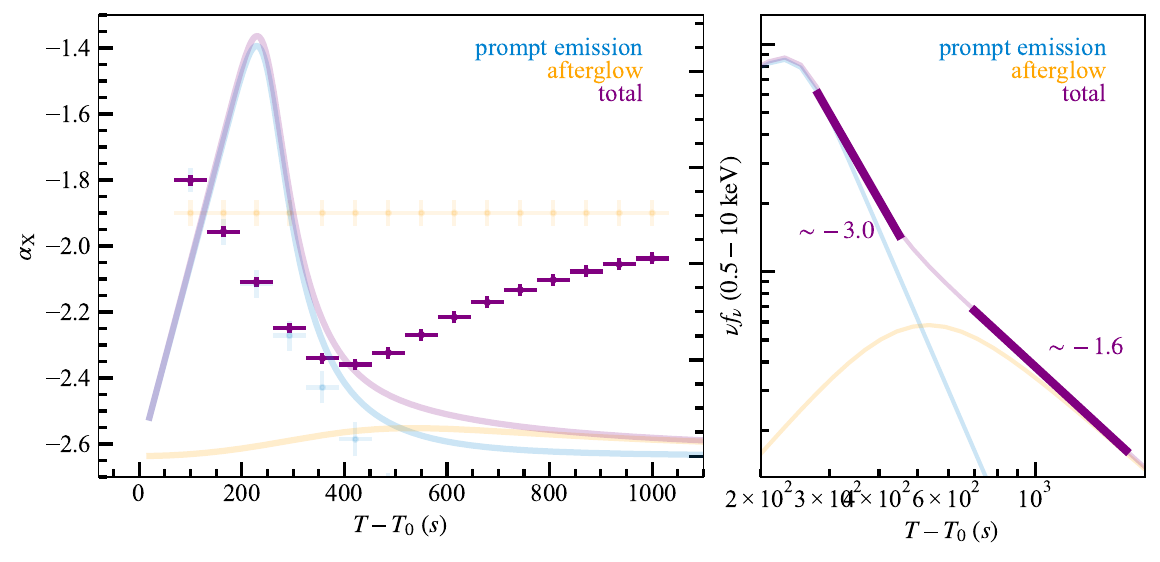}
 \includegraphics[width = 0.45\textwidth]{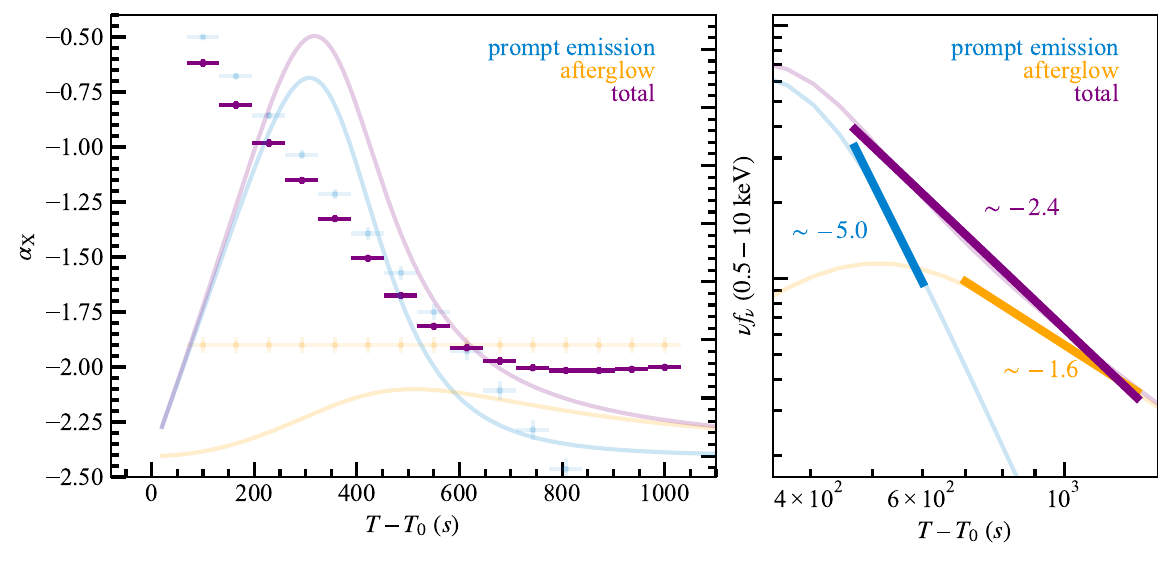}
 
  \caption{The spectral index evolutions and light curves for four scenarios of the transition from the prompt emission to the afterglow in the long-duration fast X-ray transients. The purple, blue, and yellow dots represent the spectral indices of the total, prompt, and afterglow emission spectra. The colored curves demonstrate the flux evolution in linear and logarithmic scale. From top to bottom: case I, case II, case III, and case IV.}
 \label{fig:illus}
\end{figure}

\item \textit{Case III}: Spectral hardening with a temporal break. The afterglow dominates at later times, after the prompt emission has already softened and faded. Spectral hardening is observed in the long-duration decaying tail of the continuous X-ray emission, with the flux exhibiting two distinct phases: an initial steep decay followed by a transition to a normal decay slope. This behavior is observed in this work.

\item \textit{Case IV}: Smooth spectral and temporal decay. The prompt and afterglow emissions are superimposed and have comparable intensities. At the point when the afterglow begins to dominate, the prompt emission still maintains a harder spectral index than the afterglow. Neither a significant spectral hardening nor a clear temporal break is observed. The transition appears continuous in both the light curve and spectral evolution, making it difficult to confidently identify the afterglow component or to confirm the GRB-dominated nature of the fast X-ray transient.

\end{itemize}

In summary, this study identifies a distinct spectral evolution pattern in fast X-ray transients, where the spectral index $\alpha_{\rm X}$ transitions from softening to hardening, eventually reaching a plateau. This pattern directly signals the emergence of the afterglow following the prompt emission phase. Importantly, it provides a reliable criterion for distinguishing a specific subgroup of GRBs among fast X-ray transients, even in the absence of gamma-ray counterparts or clear temporal features in the X-ray flux.

\begin{acknowledgments}
We are grateful to Xinglong-2.16 m telescope team for the helpful discussions on the redshift determination. 
We are thankful for the support of the YNAO staff at Lijiang Observatory.
We acknowledge the support by the National Key Research and Development Programs of China (2022YFF0711404, 2022SKA0130102, and 2021YFA0718500), the National SKA Program of China (2022SKA0130100), the National Natural Science Foundation of China (grant Nos. 11833003, U2038105, U1831135, 12121003, 12393811, and 13001106), the science research grants from the China Manned Space Project with No. CMS-CSST-2021-B11, and the Fundamental Research Funds for the Central Universities. 
This work is based on data obtained with the Einstein Probe, a space mission supported by the Strategic Priority Program on Space Science of the Chinese Academy of Sciences, in collaboration with ESA, MPE, and CNES (grant No. XDA15310000, No. XDA15052100). 
Mephisto is developed at and operated by the South-Western Institute for Astronomy Research of Yunnan University (SWIFAR-YNU), funded by the ``Yunnan University Development Plan for WorldClass University'' and ``Yunnan University Development Plan for World-Class Astronomy Discipline.'' The authors from YNU acknowledge support from the “Science \& Technology Champion Project'' (202005AB160002) and from two ``Team Projects,'' the ``Innovation Team'' (202105AE160021) and the ``Top Team'' (202305AT350002), all funded by the ``Yunnan Revitalization Talent Support Program''. This work is also supported by the National Key Research and Development Program of China (2024YFA1611603) and the “Yunnan Provincial Key Laboratory of Survey Science” with project No. 202449CE340002. Y.F. is supported by the Yunnan Fundamental Research Projects (202301AU070006). 
X.W. is supported by NSFC (12288102, 12033003) and the New Cornerstone Science Foundation through the XPLORER PRIZE. Members of TNOT acknowledge financial support from the Natural Science Foundation of Xinjiang Uygur Autonomous Region under No. 2024D01D32 and the Tianshan Talent Training Program grants 2023TSYCLJ0053 and 2023TSYCCX0101, the Central Guidance for Local Science and Technology Development Fund under No. ZYYD2025QY27, and the National Natural Science Foundation of China NSFC 12433007. 
The SYSU 80cm infrared telescope is operated and managed by the Department of Astronomy, Sun Yat-sen University. 
J.M. has the financial support of the National Key R\&D Program of China (2023YFE0101200), Natural Science Foundation of China 12393813, and the Yunnan Revitalization Talent Support Program (YunLing Scholar Project). 
A.J.C.T. acknowledges support from the Spanish Ministry project PID2023-151905OB-I00 and Junta de Andaluc\'ia grant P20\_010168 and from the Severo Ochoa grant CEX2021-001131-S funded by MCIN/AEI/10.13039/501100011033. The 60/90 cm Schmidt telescope at the Xinglong station is operated by the WSGP group at NAOC and supported by the CAS Special Funds for Observatory and Facility Development and the Zhengjia Enterprise Group. W.L., H.Z., Y.X., W.G., L.F., and N.L. are supported by NSFC (12120101003 and 12373010), the National Key R\&D Program (2022YFA1602902 and 2023YFA1607804), and the Strategic Priority Research Program of CAS (XDB0550100 and XDB0550000).
D.B.M. is funded by the European Union (ERC, HEAVYMETAL, 101071865). Views and opinions expressed are, however, those of the authors only and do not necessarily reflect those of the European Union or the European Research Council. Neither the European Union nor the granting authority can be held responsible for them. The Cosmic Dawn Center (DAWN) is funded by the Danish National Research Foundation under grant DNRF140.
\end{acknowledgments}

\appendix
\restartappendixnumbering

\section{Spectral Fitting Results} \label{sec:specfits}

Tables \ref{tab:spec_fit1} and \ref{tab:spec_fit2} in Appendix A summarize the best-fitting model parameters and corresponding statistics for each time slice of the time-integrated and time-resolved spectral fittings for Fermi/GBM and EP data.

\begin{table}
\centering
\scriptsize
\caption{Spectral fitting results and corresponding fitting statistics for EP/WXT and Fermi/GBM. All errors represent the 1$\sigma$ uncertainties.}
\label{tab:spec_fit1}
\begin{tabular}{ccccccc}
\hline
\hline
\multirow{2}{*}{$t1$} & \multirow{2}{*}{$t2$} & \multicolumn{4}{c}{CPL Model}\\
\cline{3-7}
& & $\alpha$ & $E_{\rm p}$ & log$A$ & pgstat/dof & BIC\\
(s) & (s) & & (keV) & ($\rm{photons~cm^{-2}~s^{-1}~keV^{-1}}$) & &\\
\hline
-4.00 & 113.00 & ${-1.15}_{-0.02}^{+0.02}$ & ${55.34}_{-0.51}^{+0.64}$ & ${-1.54}_{-0.02}^{+0.01}$ & 749.19/219 & 765.4\\
\hline
-1.00$^*$ & 15.00$^*$ & ${-0.93}_{-0.04}^{+0.05}$ & ${50.93}_{-1.62}^{+1.55}$ & ${-1.45}_{-0.04}^{+0.05}$ & 218.15/187 & 233.89\\
15.00 & 23.00 & ${-0.97}_{-0.04}^{+0.04}$ & ${76.38}_{-1.57}^{+1.78}$ & ${-1.16}_{-0.03}^{+0.03}$ & 277.31/213 & 293.44\\
23.00 & 31.00 & ${-0.87}_{-0.06}^{+0.05}$ & ${54.20}_{-1.11}^{+1.13}$ & ${-1.11}_{-0.06}^{+0.04}$ & 225.80/186 & 241.52\\
31.00 & 46.00 & ${-0.73}_{-0.06}^{+0.05}$ & ${50.58}_{-0.81}^{+0.94}$ & ${-1.07}_{-0.05}^{+0.05}$ & 248.63/182 & 264.29\\
46.00 & 54.00 & ${-1.24}_{-0.06}^{+0.05}$ & ${73.28}_{-2.32}^{+3.28}$ & ${-1.50}_{-0.05}^{+0.04}$ & 228.99/203 & 244.98\\
54.00 & 71.00 & ${-1.39}_{-0.04}^{+0.06}$ & ${55.98}_{-1.90}^{+1.83}$ & ${-1.75}_{-0.04}^{+0.05}$ & 246.00/187 & 261.74\\
71.00 & 88.00 & ${-1.23}_{-0.06}^{+0.07}$ & ${49.77}_{-1.58}^{+1.51}$ & ${-1.66}_{-0.05}^{+0.06}$ & 238.42/175 & 253.97\\
88.00 & 113.00 & ${-1.42}_{-0.10}^{+0.09}$ & ${29.65}_{-1.85}^{+1.76}$ & ${-1.87}_{-0.10}^{+0.08}$ & 202.69/143 & 217.64\\
\hline
-4.00 & 2.58 & ${-1.28}_{-0.16}^{+0.11}$ & ${127.94}_{-16.25}^{+47.05}$ & ${-2.17}_{-0.12}^{+0.07}$ & 59.36/82 & 72.68\\
2.58 & 7.00 & ${-0.73}_{-0.12}^{+0.16}$ & ${50.93}_{-2.29}^{+2.03}$ & ${-1.21}_{-0.10}^{+0.14}$ & 136.65/135 & 151.43\\
7.00 & 13.00 & ${-0.74}_{-0.20}^{+0.21}$ & ${34.28}_{-1.92}^{+1.37}$ & ${-1.18}_{-0.18}^{+0.21}$ & 109.43/113 & 123.69\\
13.00 & 17.00 & ${-1.13}_{-0.01}^{+0.12}$ & ${58.61}_{-2.51}^{+2.62}$ & ${-1.51}_{-0.01}^{+0.10}$ & 149.80/146 & 164.62\\
17.00 & 18.40 & ${-1.03}_{-0.10}^{+0.08}$ & ${86.50}_{-3.89}^{+6.83}$ & ${-1.25}_{-0.08}^{+0.06}$ & 139.14/169 & 154.58\\
18.40 & 20.00 & ${-0.82}_{-0.09}^{+0.07}$ & ${79.07}_{-2.68}^{+4.11}$ & ${-1.01}_{-0.07}^{+0.06}$ & 185.96/183 & 201.64\\
20.00 & 21.11 & ${-1.03}_{-0.11}^{+0.08}$ & ${77.98}_{-3.51}^{+5.39}$ & ${-1.13}_{-0.09}^{+0.06}$ & 172.36/172 & 187.85\\
21.11 & 23.00 & ${-0.83}_{-0.08}^{+0.07}$ & ${71.78}_{-1.96}^{+2.69}$ & ${-0.95}_{-0.06}^{+0.06}$ & 210.18/189 & 225.95\\
23.00 & 25.00 & ${-0.90}_{-0.09}^{+0.10}$ & ${57.68}_{-1.83}^{+2.30}$ & ${-1.05}_{-0.07}^{+0.08}$ & 190.75/163 & 206.08\\
25.00 & 27.00 & ${-0.71}_{-0.13}^{+0.13}$ & ${45.19}_{-1.73}^{+1.70}$ & ${-0.89}_{-0.11}^{+0.12}$ & 130.02/143 & 144.97\\
27.00 & 29.23 & ${-0.90}_{-0.13}^{+0.11}$ & ${56.62}_{-2.42}^{+3.08}$ & ${-1.22}_{-0.11}^{+0.09}$ & 124.41/138 & 139.25\\
29.23 & 32.00 & ${-0.77}_{-0.10}^{+0.13}$ & ${54.33}_{-2.09}^{+2.04}$ & ${-1.09}_{-0.08}^{+0.10}$ & 126.66/149 & 141.74\\
32.00 & 34.83 & ${-0.73}_{-0.18}^{+0.14}$ & ${44.87}_{-1.72}^{+2.11}$ & ${-1.07}_{-0.16}^{+0.12}$ & 115.43/128 & 130.05\\
34.83 & 39.00 & ${-0.51}_{-0.13}^{+0.12}$ & ${46.67}_{-1.17}^{+1.75}$ & ${-0.88}_{-0.12}^{+0.10}$ & 179.56/143 & 194.51\\
39.00 & 41.24 & ${-0.85}_{-0.11}^{+0.14}$ & ${61.94}_{-2.79}^{+3.22}$ & ${-1.21}_{-0.09}^{+0.11}$ & 123.80/135 & 138.58\\
41.24 & 43.52 & ${-0.77}_{-0.12}^{+0.15}$ & ${49.66}_{-2.24}^{+1.86}$ & ${-1.05}_{-0.10}^{+0.13}$ & 158.07/133 & 172.81\\
43.52 & 46.00 & ${-0.86}_{-0.14}^{+0.13}$ & ${54.33}_{-2.68}^{+2.69}$ & ${-1.19}_{-0.12}^{+0.11}$ & 131.53/137 & 146.36\\
46.00 & 50.00 & ${-1.08}_{-0.10}^{+0.09}$ & ${55.34}_{-2.61}^{+2.34}$ & ${-1.39}_{-0.08}^{+0.08}$ & 135.98/158 & 151.22\\
50.00 & 52.00 & ${-1.23}_{-0.06}^{+0.08}$ & ${99.08}_{-6.40}^{+6.84}$ & ${-1.43}_{-0.05}^{+0.06}$ & 190.72/192 & 206.53\\
52.00 & 54.00 & ${-1.33}_{-0.11}^{+0.09}$ & ${79.62}_{-5.31}^{+8.90}$ & ${-1.57}_{-0.09}^{+0.07}$ & 148.23/149 & 163.3\\
54.00 & 58.00 & ${-1.24}_{-0.08}^{+0.08}$ & ${68.08}_{-3.36}^{+3.54}$ & ${-1.51}_{-0.06}^{+0.06}$ & 215.77/173 & 231.28\\
58.00 & 60.84 & ${-1.25}_{-0.13}^{+0.12}$ & ${51.64}_{-3.45}^{+3.69}$ & ${-1.57}_{-0.11}^{+0.11}$ & 128.75/124 & 143.28\\
60.84 & 64.00 & ${-1.44}_{-0.12}^{+0.12}$ & ${56.89}_{-5.36}^{+4.77}$ & ${-1.78}_{-0.10}^{+0.10}$ & 125.99/120 & 140.43\\
64.00 & 71.00 & ${-1.54}_{-0.11}^{+0.13}$ & ${41.88}_{-4.04}^{+3.62}$ & ${-2.00}_{-0.10}^{+0.11}$ & 101.99/126 & 116.57\\
71.00 & 79.00 & ${-1.43}_{-0.17}^{+0.14}$ & ${38.11}_{-4.30}^{+3.68}$ & ${-2.00}_{-0.16}^{+0.13}$ & 92.43/98 & 106.28\\
79.00 & 82.52 & ${-1.17}_{-0.13}^{+0.13}$ & ${50.70}_{-3.17}^{+2.88}$ & ${-1.58}_{-0.12}^{+0.12}$ & 136.83/117 & 151.19\\
82.52 & 86.00 & ${-1.10}_{-0.09}^{+0.08}$ & ${56.10}_{-2.77}^{+2.24}$ & ${-1.37}_{-0.08}^{+0.07}$ & 157.91/157 & 173.13\\
86.00 & 89.18 & ${-1.08}_{-0.14}^{+0.12}$ & ${46.45}_{-2.50}^{+2.87}$ & ${-1.43}_{-0.12}^{+0.11}$ & 148.23/123 & 162.74\\
89.18 & 94.00 & ${-1.18}_{-0.16}^{+0.18}$ & ${33.19}_{-1.93}^{+2.29}$ & ${-1.55}_{-0.15}^{+0.16}$ & 114.38/110 & 128.56\\
94.00 & 98.00 & ${-1.23}_{-0.11}^{+0.13}$ & ${41.98}_{-2.35}^{+1.98}$ & ${-1.48}_{-0.10}^{+0.11}$ & 129.30/136 & 144.1\\
98.00 & 113.00 & ${-1.58}_{-0.22}^{+0.27}$ & ${16.52}_{-5.78}^{+3.71}$ & ${-2.12}_{-0.23}^{+0.28}$ & 118.68/91 & 132.31\\
\hline
\hline
\end{tabular}
\begin{tablenotes}
\scriptsize
\item * EP/WXT and Fermi/GBM joint fit.
\end{tablenotes}
\end{table}

\begin{table}
\centering
\tiny
\caption{Spectral fitting results and corresponding fitting statistics for EP/FXT. All errors represent the 1$\sigma$ uncertainties.}
\label{tab:spec_fit2}
\begin{tabular}{ccccccccccccc}
\hline
\hline
\multirow{2}{*}{$t1$} & \multirow{2}{*}{$t2$} & \multicolumn{4}{c}{PL Model} & & \multicolumn{6}{c}{SBPL model}\\
\cline{3-6}
\cline{8-13}
& & $\alpha$ & log$A$ & cstat/dof & BIC & & $\alpha$ & $E_{\rm break}$ & $\beta$ & log$A$ & cstat/dof & BIC\\
(s)& (s) & \multicolumn{3}{c}{($\rm{photons~cm^{-2}~s^{-1}~keV^{-1}}$)} & & & & (keV) & \multicolumn{3}{c}{($\rm{photons~cm^{-2}~s^{-1}~keV^{-1}}$)} & \\
\hline
130.00 & 255.00 & ${-2.59}_{-0.06}^{+0.05}$ & ${-4.73}_{-0.11}^{+0.09}$ & 240.77/248 & 251.82 & 
& ${-2.53}_{-0.12}^{+0.00}$ & ${2.77}_{-0.01}^{+4.14}$ & ${-2.87}_{-0.41}^{+0.31}$ & ${-5.13}_{-0.50}^{+0.46}$ & 239.83/246 & 261.92\\
255.00 & 1384.00 & ${-1.99}_{-0.02}^{+0.02}$ & ${-5.47}_{-0.04}^{+0.05}$ & 374.23/407 & 386.26 & 
& ${-1.94}_{-0.06}^{+0.02}$ & ${2.38}_{-1.22}^{+2.99}$ & ${-2.18}_{-0.39}^{+0.18}$ & ${-5.76}_{-0.48}^{+0.28}$ & 367.91/405 & 391.97\\
4324.00 & 7143.00 & ${-1.88}_{-0.05}^{+0.05}$ & ${-6.34}_{-0.09}^{+0.09}$ & 232.47/255 & 243.57 & 
& ${-1.74}_{-0.19}^{+0.11}$ & ${2.09}_{-1.48}^{+3.20}$ & ${-2.21}_{-0.06}^{+0.38}$ & ${-6.86}_{-0.07}^{+0.61}$ & 227.93/253 & 250.13\\
\hline
130.00 & 140.50 & ${-2.23}_{-0.11}^{+0.12}$ & ${-3.59}_{-0.20}^{+0.22}$ & 109.46/117 & 119.02 & 
& ${-2.07}_{-0.17}^{+0.19}$ & ${2.55}_{-0.36}^{+2.09}$ & ${-3.22}_{-1.29}^{+0.40}$ & ${-5.10}_{-1.66}^{+0.63}$ & 104.68/115 & 123.8\\
140.50 & 154.00 & ${-2.38}_{-0.09}^{+0.11}$ & ${-3.97}_{-0.17}^{+0.20}$ & 115.88/115 & 125.4 & 
& ${-1.98}_{-0.27}^{+0.17}$ & ${1.86}_{-0.32}^{+0.63}$ & ${-3.21}_{-0.64}^{+0.40}$ & ${-5.32}_{-1.03}^{+0.66}$ & 107.98/113 & 127.03\\
154.00 & 164.50 & ${-2.67}_{-0.18}^{+0.14}$ & ${-4.63}_{-0.33}^{+0.27}$ & 69.70/80 & 78.52 & 
& ${-3.22}_{-1.87}^{+1.07}$ & ${1.02}_{-0.47}^{+4.10}$ & ${-2.56}_{-1.07}^{+0.05}$ & ${-4.45}_{-1.55}^{+0.11}$ & 69.03/78 & 86.66\\
164.50 & 180.00 & ${-2.73}_{-0.15}^{+0.16}$ & ${-4.92}_{-0.30}^{+0.30}$ & 65.03/77 & 73.77 & 
& ${-3.02}_{-1.59}^{+0.77}$ & ${1.14}_{-0.57}^{+2.16}$ & ${-2.61}_{-0.66}^{+0.07}$ & ${-4.69}_{-1.12}^{+0.12}$ & 64.23/75 & 81.69\\
180.00 & 255.00 & ${-3.47}_{-0.17}^{+0.14}$ & ${-6.94}_{-0.33}^{+0.28}$ & 80.29/86 & 89.25 & 
& ${-3.46}_{-1.13}^{+0.94}$ & ${3.02}_{-2.46}^{+0.05}$ & ${-2.19}_{-1.42}^{+0.18}$ & ${-5.01}_{-2.19}^{+0.04}$ & 78.35/84 & 95.9\\
255.00 & 290.50 & ${-3.15}_{-0.12}^{+0.12}$ & ${-7.48}_{-0.23}^{+0.24}$ & 120.07/112 & 129.54 & 
& ${-3.50}_{-0.30}^{+0.25}$ & ${1.22}_{-0.19}^{+0.26}$ & ${-2.61}_{-0.24}^{+0.40}$ & ${-6.53}_{-0.44}^{+0.62}$ & 111.68/110 & 130.63\\
290.50 & 359.50 & ${-2.37}_{-0.10}^{+0.10}$ & ${-6.26}_{-0.20}^{+0.20}$ & 120.56/112 & 130.03 & 
& ${-3.10}_{-1.32}^{+0.53}$ & ${0.95}_{-0.30}^{+0.26}$ & ${-2.09}_{-0.27}^{+0.13}$ & ${-5.75}_{-0.50}^{+0.21}$ & 113.79/110 & 132.74\\
359.50 & 462.00 & ${-2.06}_{-0.10}^{+0.09}$ & ${-5.81}_{-0.19}^{+0.17}$ & 137.35/115 & 146.87 & 
& ${-4.42}_{-1.53}^{+1.26}$ & ${0.71}_{-0.09}^{+0.10}$ & ${-1.92}_{-0.15}^{+0.11}$ & ${-5.56}_{-0.27}^{+0.21}$ & 129.63/113 & 148.68\\
462.00 & 571.50 & ${-2.03}_{-0.10}^{+0.10}$ & ${-5.78}_{-0.19}^{+0.18}$ & 102.72/125 & 112.41 & 
& ${-5.27}_{-0.15}^{+3.30}$ & ${0.58}_{-0.01}^{+5.17}$ & ${-2.00}_{-0.87}^{+0.15}$ & ${-5.75}_{-1.08}^{+0.27}$ & 101.79/123 & 120.72\\
571.50 & 650.50 & ${-1.94}_{-0.10}^{+0.08}$ & ${-5.46}_{-0.19}^{+0.16}$ & 119.56/125 & 129.24 & 
& ${-2.62}_{-2.36}^{+0.70}$ & ${0.98}_{-0.41}^{+2.11}$ & ${-1.80}_{-0.34}^{+0.04}$ & ${-5.22}_{-0.54}^{+0.08}$ & 116.54/123 & 135.91\\
650.50 & 729.00 & ${-1.81}_{-0.09}^{+0.09}$ & ${-5.22}_{-0.16}^{+0.17}$ & 147.82/125 & 157.51 & 
& ${-1.71}_{-0.16}^{+0.13}$ & ${2.65}_{-1.13}^{+2.19}$ & ${-2.68}_{-0.45}^{+0.81}$ & ${-6.55}_{-0.46}^{+1.23}$ & 144.49/123 & 163.87\\
729.00 & 806.50 & ${-1.96}_{-0.10}^{+0.08}$ & ${-5.50}_{-0.20}^{+0.16}$ & 73.82/120 & 83.43 & 
& ${-1.89}_{-0.77}^{+0.29}$ & ${3.01}_{-2.44}^{+1.92}$ & ${-2.67}_{-0.49}^{+0.78}$ & ${-6.54}_{-0.51}^{+1.19}$ & 72.73/118 & 91.46\\
806.50 & 868.00 & ${-1.85}_{-0.10}^{+0.08}$ & ${-5.18}_{-0.20}^{+0.15}$ & 106.86/129 & 116.62 & 
& ${-1.75}_{-0.88}^{+0.22}$ & ${3.42}_{-2.86}^{+4.08}$ & ${-2.74}_{-0.13}^{+1.00}$ & ${-6.43}_{-0.17}^{+1.45}$ & 104.65/127 & 123.85\\
868.00 & 921.00 & ${-1.81}_{-0.10}^{+0.09}$ & ${-5.05}_{-0.19}^{+0.17}$ & 110.36/120 & 119.97 & 
& ${-1.65}_{-0.36}^{+0.18}$ & ${2.07}_{-1.47}^{+3.56}$ & ${-2.24}_{-0.55}^{+0.50}$ & ${-5.74}_{-0.66}^{+0.81}$ & 107.91/118 & 127.13\\
921.00 & 973.50 & ${-2.02}_{-0.08}^{+0.09}$ & ${-5.43}_{-0.17}^{+0.17}$ & 112.56/124 & 122.23 & 
& ${-1.95}_{-1.75}^{+0.10}$ & ${2.39}_{-1.82}^{+3.43}$ & ${-2.44}_{-0.15}^{+0.63}$ & ${-6.09}_{-0.11}^{+0.98}$ & 111.00/122 & 130.34\\
973.50 & 1018.50 & ${-1.97}_{-0.09}^{+0.09}$ & ${-5.28}_{-0.18}^{+0.16}$ & 122.52/122 & 132.16 & 
& ${-1.47}_{-0.67}^{+0.54}$ & ${1.13}_{-0.54}^{+2.48}$ & ${-2.20}_{-0.07}^{+0.32}$ & ${-5.67}_{-0.13}^{+0.56}$ & 119.57/120 & 138.86\\
1018.50 & 1065.50 & ${-1.89}_{-0.10}^{+0.09}$ & ${-5.13}_{-0.19}^{+0.18}$ & 123.27/120 & 132.88 & 
& ${-1.37}_{-0.26}^{+0.19}$ & ${1.58}_{-0.23}^{+0.37}$ & ${-2.55}_{-0.40}^{+0.23}$ & ${-6.22}_{-0.65}^{+0.37}$ & 109.89/118 & 129.11\\
1065.50 & 1117.50 & ${-1.92}_{-0.10}^{+0.09}$ & ${-5.24}_{-0.18}^{+0.18}$ & 117.79/122 & 127.43 & 
& ${-0.84}_{-1.10}^{+0.28}$ & ${0.77}_{-0.11}^{+3.23}$ & ${-2.02}_{-0.19}^{+0.18}$ & ${-5.44}_{-0.32}^{+0.29}$ & 114.85/120 & 133.89\\
1117.50 & 1168.50 & ${-1.93}_{-0.09}^{+0.09}$ & ${-5.24}_{-0.18}^{+0.16}$ & 109.55/126 & 119.25 & 
& ${-2.20}_{-2.24}^{+0.38}$ & ${0.89}_{-0.33}^{+5.32}$ & ${-1.86}_{-0.69}^{+0.10}$ & ${-5.13}_{-0.98}^{+0.17}$ & 108.87/124 & 127.64\\
1168.50 & 1222.50 & ${-2.02}_{-0.09}^{+0.08}$ & ${-5.46}_{-0.19}^{+0.16}$ & 112.52/121 & 122.15 & 
& ${-1.90}_{-0.36}^{+0.18}$ & ${3.24}_{-2.66}^{+1.96}$ & ${-2.71}_{-0.10}^{+0.84}$ & ${-6.44}_{-0.23}^{+1.21}$ & 111.06/119 & 129.42\\
1222.50 & 1276.50 & ${-1.76}_{-0.09}^{+0.09}$ & ${-4.95}_{-0.18}^{+0.17}$ & 139.92/126 & 149.62 & 
& ${-1.63}_{-0.13}^{+0.15}$ & ${3.55}_{-1.93}^{+0.93}$ & ${-3.35}_{-0.19}^{+1.40}$ & ${-7.19}_{-0.19}^{+1.92}$ & 134.71/124 & 153.63\\
1276.50 & 1384.00 & ${-1.91}_{-0.06}^{+0.07}$ & ${-5.28}_{-0.12}^{+0.13}$ & 173.10/183 & 183.54 & 
& ${-1.72}_{-0.14}^{+0.43}$ & ${2.23}_{-1.39}^{+0.96}$ & ${-2.49}_{-0.23}^{+0.51}$ & ${-6.17}_{-0.29}^{+0.79}$ & 167.06/181 & 187.94\\
4324.00 & 4857.50 & ${-1.88}_{-0.09}^{+0.08}$ & ${-6.18}_{-0.18}^{+0.15}$ & 113.10/121 & 122.72 & 
& ${-1.66}_{-0.15}^{+0.27}$ & ${2.02}_{-0.71}^{+0.67}$ & ${-2.71}_{-0.11}^{+0.61}$ & ${-7.53}_{-0.16}^{+1.02}$ & 104.91/119 & 124.16\\
4857.50 & 5552.50 & ${-1.85}_{-0.09}^{+0.09}$ & ${-6.24}_{-0.17}^{+0.17}$ & 105.77/122 & 115.41 & 
& ${-1.80}_{-0.52}^{+0.56}$ & ${1.75}_{-1.18}^{+3.49}$ & ${-1.97}_{-0.12}^{+0.47}$ & ${-6.44}_{-0.19}^{+0.64}$ & 105.43/120 & 123.93\\
5552.50 & 7143.00 & ${-1.90}_{-0.08}^{+0.07}$ & ${-6.49}_{-0.15}^{+0.14}$ & 139.65/170 & 149.95 & 
& ${-2.47}_{-0.97}^{+0.66}$ & ${0.91}_{-0.34}^{+4.49}$ & ${-1.75}_{-0.18}^{+0.66}$ & ${-6.24}_{-0.32}^{+0.79}$ & 137.44/168 & 158.03\\
44412.00 & 47458.00 & ${-2.02}_{-0.25}^{+0.23}$ & ${-8.11}_{-0.50}^{+0.42}$ & 15.35/19 & 24.62 & 
& ${-4.04}_{-1.24}^{+1.56}$ & ${0.95}_{-0.23}^{+0.35}$ & ${-1.60}_{-0.17}^{+0.44}$ & ${-7.40}_{-0.28}^{+0.71}$ & 10.12/17 & 25.57\\
114594.00 & 122326.00 & ${-1.53}_{-0.44}^{+0.30}$ & ${-7.70}_{-0.87}^{+0.53}$ & 12.65/7 & 19.56 & 
& ${-4.12}_{-0.73}^{+2.06}$ & ${1.11}_{-0.27}^{+0.47}$ & ${-0.70}_{-0.64}^{+0.25}$ & ${-6.40}_{-1.02}^{+0.38}$ & 7.81/5 & 19.32\\
\hline
\hline
\end{tabular}
\end{table}

\section{The Photometric and Spectroscopic Results}
\label{sec:photo}

\begin{figure}
 \centering
 \includegraphics[width = 0.90\textwidth]{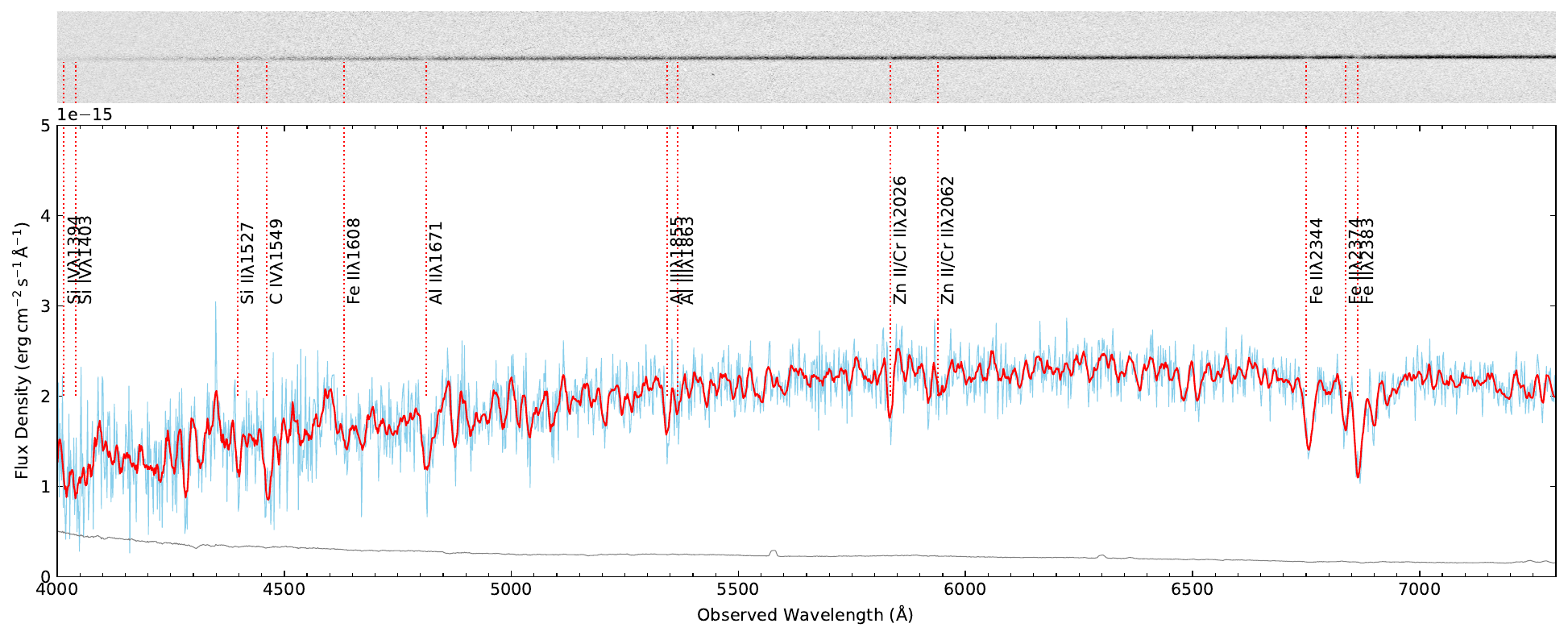}
 \caption{The optical spectrum observed by the GMG-2.4m telescope. The wavelength-calibrated 2D spectrum and the corresponding flux-calibrated 1D spectrum are shown. The 1D raw, smoothed, and error spectra are shown in sky blue, red, and gray lines, respectively. Multiple metal absorption lines at a redshift of $z$ = 1.88 are marked with vertical dashed lines. Note that the flux calibration should be treated with caution due to the absence of a standard star observation on the same night. Instead, a previously obtained standard star spectrum was used for the flux calibration.}
 \label{fig:redshift}
\end{figure}

Appendix B presents the optical and near-infrared photometry of GRB 250404A/EP250404a used in the afterglow analysis in Tables \ref{tab:photoag} and \ref{tab:photoclear}, as well as the GMG-2.4 m optical spectrum in Figure \ref{fig:redshift}.

\begin{ThreePartTable}
  \begin{TableNotes}
  \item[*] Upper limits.
  \end{TableNotes}
\begin{center}
\begin{longtable}{cccc}
  \caption{Optical and near-infrared observations of GRB 250404A/EP250404a for afterglow fitting. The magnitudes have not been corrected for foreground Galactic extinction. All errors represent the 1$\sigma$ uncertainties.}
  \label{tab:photoag} \\
  \hline
  $T-T_0~(\rm s)$ & Telescope & Band & AB Magnitude \\
  \hline
  \endfirsthead 
  \multicolumn{4}{c}{{\textbf{\tablename\ \thetable{}.} Continued}} \\
  \hline
  $T-T_0~(\rm s)$ & Telescope & Band & AB Magnitude \\
  \hline
  \endhead
  \insertTableNotes
  \endlastfoot 
734 & Mephisto & $g$ & 15.72 $\pm$ 0.01 \\
734 & Mephisto & $i$ & 14.58 $\pm$ 0.00 \\
734 & Mephisto & $u$ & 18.65 $\pm$ 0.08 \\
794 & Mephisto & $g$ & 15.53 $\pm$ 0.01 \\
794 & Mephisto & $i$ & 14.30 $\pm$ 0.00 \\
854 & Mephisto & $g$ & 15.37 $\pm$ 0.01 \\
914 & Mephisto & $g$ & 15.24 $\pm$ 0.01 \\
914 & Mephisto & $i$ & 14.15 $\pm$ 0.00 \\
914 & Mephisto & $u$ & 18.37 $\pm$ 0.07 \\
974 & Mephisto & $g$ & 15.23 $\pm$ 0.01 \\
1034 & Mephisto & $g$ & 15.24 $\pm$ 0.01 \\
1034 & Mephisto & $i$ & 14.31 $\pm$ 0.00 \\
1154 & Mephisto & $r$ & 14.83 $\pm$ 0.01 \\
1154 & Mephisto & $z$ & 14.04 $\pm$ 0.01 \\
1154 & Mephisto & $v$ & 16.91 $\pm$ 0.02 \\
1214 & TNOT & $rp$ & 14.98 $\pm$ 0.04 \\
1214 & Mephisto & $r$ & 14.91 $\pm$ 0.01 \\
1214 & Mephisto & $z$ & 14.20 $\pm$ 0.01 \\
1257 & ALT100C & $r$ & 14.91 $\pm$ 0.01 \\
1274 & Mephisto & $r$ & 15.02 $\pm$ 0.01 \\
1322 & ALT100C & $r$ & 15.03 $\pm$ 0.01 \\
1331 & TNOT & $rp$ & 15.19 $\pm$ 0.05 \\
1334 & Mephisto & $r$ & 15.19 $\pm$ 0.01 \\
1334 & Mephisto & $z$ & 14.45 $\pm$ 0.01 \\
1334 & Mephisto & $v$ & 17.28 $\pm$ 0.03 \\
1388 & ALT100C & $r$ & 15.19 $\pm$ 0.01 \\
1394 & Mephisto & $r$ & 15.31 $\pm$ 0.01 \\
1443 & TNOT & $rp$ & 15.40 $\pm$ 0.05 \\
1454 & Mephisto & $r$ & 15.38 $\pm$ 0.01 \\
1454 & Mephisto & $z$ & 14.58 $\pm$ 0.01 \\
1454 & ALT100C & $r$ & 15.37 $\pm$ 0.01 \\
1519 & ALT100C & $r$ & 15.39 $\pm$ 0.01 \\
1555 & TNOT & $rp$ & 15.50 $\pm$ 0.04 \\
1595 & ALT100C & $i$ & 15.04 $\pm$ 0.01 \\
1634 & Mephisto & $r$ & 15.55 $\pm$ 0.01 \\
1634 & Mephisto & $z$ & 14.79 $\pm$ 0.01 \\
1634 & Mephisto & $v$ & 17.51 $\pm$ 0.04 \\
1661 & ALT100C & $i$ & 15.10 $\pm$ 0.01 \\
1667 & TNOT & $rp$ & 15.62 $\pm$ 0.05 \\
1694 & Mephisto & $r$ & 15.61 $\pm$ 0.01 \\
1726 & ALT100C & $i$ & 15.18 $\pm$ 0.01 \\
1754 & Mephisto & $r$ & 15.63 $\pm$ 0.01 \\
1754 & Mephisto & $z$ & 14.84 $\pm$ 0.01 \\
1780 & TNOT & $rp$ & 15.68 $\pm$ 0.05 \\
1814 & Mephisto & $r$ & 15.69 $\pm$ 0.01 \\
1814 & Mephisto & $z$ & 14.94 $\pm$ 0.01 \\
1814 & Mephisto & $v$ & 17.68 $\pm$ 0.04 \\
1874 & Mephisto & $r$ & 15.75 $\pm$ 0.01 \\
1885 & ALT100C & $r$ & 15.75 $\pm$ 0.01 \\
1892 & TNOT & $rp$ & 15.81 $\pm$ 0.06 \\
1934 & Mephisto & $r$ & 15.80 $\pm$ 0.01 \\
1934 & Mephisto & $z$ & 15.04 $\pm$ 0.01 \\
1951 & ALT100C & $r$ & 15.80 $\pm$ 0.01 \\
2005 & TNOT & $rp$ & 15.88 $\pm$ 0.04 \\
2017 & ALT100C & $r$ & 15.83 $\pm$ 0.01 \\
2054 & Mephisto & $r$ & 15.93 $\pm$ 0.02 \\
2054 & Mephisto & $z$ & 15.16 $\pm$ 0.01 \\
2054 & Mephisto & $v$ & 17.98 $\pm$ 0.05 \\
2082 & ALT100C & $r$ & 15.91 $\pm$ 0.01 \\
2114 & Mephisto & $r$ & 16.01 $\pm$ 0.02 \\
2118 & TNOT & $rp$ & 16.04 $\pm$ 0.05 \\
2148 & ALT100C & $r$ & 16.01 $\pm$ 0.01 \\
2174 & Mephisto & $r$ & 16.05 $\pm$ 0.02 \\
2174 & Mephisto & $z$ & 15.28 $\pm$ 0.02 \\
2224 & ALT100C & $i$ & 15.66 $\pm$ 0.01 \\
2229 & TNOT & $rp$ & 16.11 $\pm$ 0.04 \\
2234 & Mephisto & $r$ & 16.10 $\pm$ 0.02 \\
2234 & Mephisto & $v$ & 18.07 $\pm$ 0.06 \\
2290 & ALT100C & $i$ & 15.68 $\pm$ 0.01 \\
2294 & Mephisto & $z$ & 15.39 $\pm$ 0.02 \\
2342 & TNOT & $rp$ & 16.19 $\pm$ 0.05 \\
2354 & Mephisto & $r$ & 16.14 $\pm$ 0.02 \\
2354 & Mephisto & $z$ & 15.44 $\pm$ 0.02 \\
2355 & ALT100C & $i$ & 15.75 $\pm$ 0.01 \\
2414 & Mephisto & $r$ & 16.23 $\pm$ 0.02 \\
2421 & ALT100C & $i$ & 15.77 $\pm$ 0.01 \\
2455 & TNOT & $rp$ & 16.34 $\pm$ 0.06 \\
2487 & ALT100C & $i$ & 15.85 $\pm$ 0.01 \\
2562 & ALT100C & $z$ & 15.66 $\pm$ 0.03 \\
2567 & TNOT & $rp$ & 16.45 $\pm$ 0.06 \\
2594 & Mephisto & $g$ & 16.84 $\pm$ 0.03 \\
2594 & Mephisto & $i$ & 15.88 $\pm$ 0.01 \\
2594 & Mephisto & $u$ & 20.05 $\pm$ 0.29 \\
2628 & ALT100C & $z$ & 15.69 $\pm$ 0.03 \\
2654 & Mephisto & $g$ & 16.96 $\pm$ 0.04 \\
2680 & TNOT & $rp$ & 16.50 $\pm$ 0.05 \\
2694 & ALT100C & $z$ & 15.83 $\pm$ 0.03 \\
2714 & Mephisto & $g$ & 17.07 $\pm$ 0.04 \\
2714 & Mephisto & $i$ & 15.97 $\pm$ 0.01 \\
2759 & ALT100C & $z$ & 15.74 $\pm$ 0.03 \\
2774 & Mephisto & $g$ & 17.03 $\pm$ 0.04 \\
2774 & Mephisto & $u$ & 19.45 $\pm$ 0.25 \\
2793 & TNOT & $rp$ & 16.64 $\pm$ 0.05 \\
2825 & ALT100C & $z$ & 15.83 $\pm$ 0.03 \\
2834 & Mephisto & $i$ & 16.04 $\pm$ 0.01 \\
2894 & Mephisto & $g$ & 17.17 $\pm$ 0.04 \\
2894 & Mephisto & $i$ & 16.05 $\pm$ 0.01 \\
2904 & ALT100C & $g$ & 17.42 $\pm$ 0.04 \\
2905 & TNOT & $rp$ & 16.63 $\pm$ 0.05 \\
2954 & Mephisto & $g$ & 17.10 $\pm$ 0.04 \\
2970 & ALT100C & $g$ & 17.43 $\pm$ 0.05 \\
3018 & TNOT & $rp$ & 16.74 $\pm$ 0.05 \\
3036 & ALT100C & $g$ & 17.47 $\pm$ 0.04 \\
3074 & Mephisto & $r$ & 16.80 $\pm$ 0.03 \\
3074 & Mephisto & $z$ & 15.96 $\pm$ 0.02 \\
3074 & Mephisto & $v$ & 18.71 $\pm$ 0.10 \\
3101 & ALT100C & $g$ & 17.50 $\pm$ 0.04 \\
3130 & TNOT & $rp$ & 16.84 $\pm$ 0.06 \\
3134 & Mephisto & $r$ & 16.79 $\pm$ 0.03 \\
3134 & Mephisto & $z$ & 16.13 $\pm$ 0.03 \\
3167 & ALT100C & $g$ & 17.52 $\pm$ 0.04 \\
3194 & Mephisto & $r$ & 16.77 $\pm$ 0.03 \\
3254 & Mephisto & $r$ & 16.84 $\pm$ 0.03 \\
3254 & Mephisto & $z$ & 16.04 $\pm$ 0.03 \\
3254 & Mephisto & $v$ & 18.82 $\pm$ 0.10 \\
3303 & ALT100C & $r$ & 16.88 $\pm$ 0.01 \\
3314 & Mephisto & $r$ & 16.86 $\pm$ 0.03 \\
3314 & Mephisto & $z$ & 16.17 $\pm$ 0.03 \\
3374 & Mephisto & $r$ & 16.91 $\pm$ 0.03 \\
3489 & ALT100C & $r$ & 16.96 $\pm$ 0.01 \\
3494 & Mephisto & $r$ & 16.97 $\pm$ 0.04 \\
3494 & Mephisto & $z$ & 16.26 $\pm$ 0.03 \\
3494 & Mephisto & $v$ & 18.56 $\pm$ 0.13 \\
3552 & ALT50D & $r$ & 16.92 $\pm$ 0.04 \\
3554 & Mephisto & $r$ & 16.98 $\pm$ 0.04 \\
3614 & Mephisto & $r$ & 17.07 $\pm$ 0.04 \\
3614 & Mephisto & $z$ & 16.28 $\pm$ 0.03 \\
3673 & ALT50D & $r$ & 17.03 $\pm$ 0.04 \\
3674 & ALT100C & $r$ & 17.03 $\pm$ 0.01 \\
3734 & Mephisto & $r$ & 16.96 $\pm$ 0.04 \\
3734 & Mephisto & $z$ & 16.40 $\pm$ 0.04 \\
3734 & Mephisto & $v$ & 19.15 $\pm$ 0.15 \\
3794 & Mephisto & $r$ & 17.22 $\pm$ 0.04 \\
3794 & Mephisto & $z$ & 16.40 $\pm$ 0.04 \\
3795 & ALT50D & $r$ & 17.11 $\pm$ 0.05 \\
3854 & Mephisto & $r$ & 17.13 $\pm$ 0.04 \\
3860 & ALT100C & $r$ & 17.14 $\pm$ 0.01 \\
3916 & ALT50D & $r$ & 17.13 $\pm$ 0.06 \\
3974 & Mephisto & $g$ & 17.65 $\pm$ 0.06 \\
3974 & Mephisto & $i$ & 16.69 $\pm$ 0.02 \\
3974 & Mephisto & $u$ & 20.41 $\pm$ 0.34 \\
4021 & SYSU 80cm & $J$ & 16.21 $\pm$ 0.08 \\
4034 & Mephisto & $g$ & 17.67 $\pm$ 0.06 \\
4037 & ALT50D & $r$ & 17.24 $\pm$ 0.07 \\
4046 & ALT100C & $r$ & 17.25 $\pm$ 0.02 \\
4094 & Mephisto & $g$ & 17.73 $\pm$ 0.07 \\
4094 & Mephisto & $i$ & 16.74 $\pm$ 0.03 \\
4214 & Mephisto & $g$ & 17.78 $\pm$ 0.07 \\
4214 & Mephisto & $i$ & 16.73 $\pm$ 0.03 \\
4216 & ALT50D & $r$ & 17.32 $\pm$ 0.05 \\
4221 & SYSU 80cm & $J$ & 16.19 $\pm$ 0.09 \\
4242 & ALT100C & $i$ & 16.95 $\pm$ 0.02 \\
4274 & Mephisto & $g$ & 17.87 $\pm$ 0.08 \\
4274 & Mephisto & $i$ & 16.85 $\pm$ 0.03 \\
4334 & Mephisto & $g$ & 17.80 $\pm$ 0.08 \\
4420 & ALT50D & $i$ & 16.83 $\pm$ 0.07 \\
4427 & ALT100C & $i$ & 17.00 $\pm$ 0.02 \\
4441 & SYSU 80cm & $J$ & 16.38 $\pm$ 0.10 \\
4454 & Mephisto & $r$ & 17.48 $\pm$ 0.03 \\
4454 & Mephisto & $z$ & 16.70 $\pm$ 0.04 \\
4454 & Mephisto & $v$ & 19.06 $\pm$ 0.12 \\
4613 & ALT100C & $i$ & 17.11 $\pm$ 0.03 \\
4624 & ALT50D & $r$ & 17.64 $\pm$ 0.13 \\
4634 & Mephisto & $r$ & 17.52 $\pm$ 0.04 \\
4634 & Mephisto & $z$ & 16.67 $\pm$ 0.04 \\
4641 & SYSU 80cm & $J$ & 16.67 $\pm$ 0.11 \\
4799 & ALT100C & $i$ & 17.09 $\pm$ 0.03 \\
4814 & Mephisto & $r$ & 17.58 $\pm$ 0.04 \\
4814 & Mephisto & $z$ & 16.76 $\pm$ 0.04 \\
4814 & Mephisto & $v$ & 19.53 $\pm$ 0.20 \\
4828 & ALT50D & $i$ & 17.02 $\pm$ 0.13 \\
4833 & TNOT & $rp$ & 17.57 $\pm$ 0.07 \\
4921 & SYSU 80cm & $J$ & 16.74 $\pm$ 0.10 \\
4947 & TNOT & $ip$ & 17.37 $\pm$ 0.14 \\
4984 & ALT100C & $i$ & 17.11 $\pm$ 0.02 \\
4994 & Mephisto & $r$ & 17.62 $\pm$ 0.04 \\
4994 & Mephisto & $z$ & 16.79 $\pm$ 0.04 \\
5032 & ALT50D & $r$ & 17.61 $\pm$ 0.08 \\
5064 & TNOT & $gp$ & 18.36 $\pm$ 0.11 \\
5174 & Mephisto & $g$ & 17.88 $\pm$ 0.06 \\
5174 & Mephisto & $i$ & 17.03 $\pm$ 0.03 \\
5182 & ALT100C & $g$ & 18.41 $\pm$ 0.04 \\
5184 & TNOT & $rp$ & 17.58 $\pm$ 0.07 \\
5236 & ALT50D & $i$ & 17.12 $\pm$ 0.10 \\
5299 & TNOT & $ip$ & 17.34 $\pm$ 0.14 \\
5354 & Mephisto & $g$ & 18.23 $\pm$ 0.06 \\
5354 & Mephisto & $i$ & 17.15 $\pm$ 0.03 \\
5368 & ALT100C & $g$ & 18.51 $\pm$ 0.05 \\
5417 & TNOT & $gp$ & 18.47 $\pm$ 0.11 \\
5440 & ALT50D & $r$ & 17.56 $\pm$ 0.10 \\
5534 & Mephisto & $g$ & 18.40 $\pm$ 0.07 \\
5534 & Mephisto & $i$ & 17.26 $\pm$ 0.03 \\
5534 & Mephisto & $u$ & 20.38 $\pm$ 0.21 \\
5538 & TNOT & $rp$ & 17.88 $\pm$ 0.08 \\
5553 & ALT100C & $g$ & 18.43 $\pm$ 0.06 \\
5621 & SYSU 80cm & $J$ & 16.96 $\pm$ 0.12 \\
5652 & TNOT & $ip$ & 17.28 $\pm$ 0.14 \\
5714 & Mephisto & $g$ & 18.25 $\pm$ 0.07 \\
5714 & Mephisto & $i$ & 17.19 $\pm$ 0.03 \\
5739 & ALT100C & $g$ & 18.63 $\pm$ 0.07 \\
5770 & TNOT & $gp$ & 18.66 $\pm$ 0.13 \\
5848 & ALT50D & $r$ & 18.20 $\pm$ 0.22 \\
5891 & TNOT & $rp$ & 17.73 $\pm$ 0.07 \\
5894 & Mephisto & $r$ & 17.88 $\pm$ 0.05 \\
5894 & Mephisto & $z$ & 17.03 $\pm$ 0.05 \\
5894 & Mephisto & $v$ & 20.19 $\pm$ 0.31 \\
5925 & ALT100C & $g$ & 18.58 $\pm$ 0.06 \\
6005 & TNOT & $ip$ & 17.60 $\pm$ 0.15 \\
6021 & SYSU 80cm & $J$ & 16.90 $\pm$ 0.12 \\
6052 & ALT50D & $i$ & 17.57 $\pm$ 0.16 \\
6074 & Mephisto & $r$ & 17.94 $\pm$ 0.05 \\
6074 & Mephisto & $z$ & 17.09 $\pm$ 0.05 \\
6123 & TNOT & $gp$ & 18.55 $\pm$ 0.15 \\
6124 & ALT100C & $z$ & 17.15 $\pm$ 0.04 \\
6254 & Mephisto & $r$ & 18.11 $\pm$ 0.06 \\
6254 & Mephisto & $z$ & 17.23 $\pm$ 0.06 \\
6254 & Mephisto & $v$ & 20.06 $\pm$ 0.22 \\
6257 & ALT50D & $r$ & 18.03 $\pm$ 0.17 \\
6309 & ALT100C & $z$ & 17.22 $\pm$ 0.05 \\
6431 & SYSU 80cm & $J$ & 16.90 $\pm$ 0.12 \\
6434 & Mephisto & $r$ & 18.11 $\pm$ 0.06 \\
6434 & Mephisto & $z$ & 17.23 $\pm$ 0.06 \\
6474 & TNOT & $ip$ & 17.57 $\pm$ 0.16 \\
6495 & ALT100C & $z$ & 17.29 $\pm$ 0.04 \\
6680 & ALT100C & $z$ & 17.47 $\pm$ 0.04 \\
6701 & TNOT & $gp$ & 18.75 $\pm$ 0.15 \\
6851 & SYSU 80cm & $J$ & 17.20 $\pm$ 0.15 \\
6866 & ALT100C & $z$ & 17.42 $\pm$ 0.04 \\
6869 & ALT50D & $i$ & 17.71 $\pm$ 0.15 \\
6931 & TNOT & $rp$ & 18.13 $\pm$ 0.09 \\
7034 & Mephisto & $g$ & 18.70 $\pm$ 0.12 \\
7034 & Mephisto & $i$ & 17.64 $\pm$ 0.06 \\
7034 & Mephisto & $u$ & 21.14 $\pm$ 0.54 \\
7073 & ALT50D & $r$ & 18.16 $\pm$ 0.10 \\
7123 & ALT100C & $r$ & 18.20 $\pm$ 0.02 \\
7154 & Mephisto & $g$ & 18.59 $\pm$ 0.11 \\
7154 & Mephisto & $i$ & 17.62 $\pm$ 0.05 \\
7161 & TNOT & $ip$ & 17.92 $\pm$ 0.18 \\
7214 & Mephisto & $g$ & 18.66 $\pm$ 0.15 \\
7214 & Mephisto & $i$ & 17.64 $\pm$ 0.06 \\
7391 & TNOT & $gp$ & 19.23 $\pm$ 0.20 \\
7394 & Mephisto & $g$ & 18.54 $\pm$ 0.12 \\
7394 & Mephisto & $i$ & 17.67 $\pm$ 0.06 \\
7429 & ALT100C & $r$ & 18.30 $\pm$ 0.03 \\
7454 & Mephisto & $g$ & 18.58 $\pm$ 0.12 \\
7454 & Mephisto & $i$ & 17.72 $\pm$ 0.06 \\
7481 & ALT50D & $r$ & 18.19 $\pm$ 0.12 \\
7574 & Mephisto & $g$ & 18.52 $\pm$ 0.11 \\
7574 & Mephisto & $i$ & 17.72 $\pm$ 0.06 \\
7611 & SYSU 80cm & $J$ & 17.12 $\pm$ 0.11 \\
7620 & TNOT & $rp$ & 18.26 $\pm$ 0.10 \\
7694 & Mephisto & $g$ & 18.64 $\pm$ 0.13 \\
7694 & Mephisto & $i$ & 17.82 $\pm$ 0.06 \\
7735 & ALT100C & $r$ & 18.25 $\pm$ 0.03 \\
7851 & TNOT & $ip$ & 18.09 $\pm$ 0.19 \\
7889 & ALT50D & $r$ & 18.33 $\pm$ 0.13 \\
7889 & ALT50D & $i$ & 18.14 $\pm$ 0.16 \\
7994 & Mephisto & $r$ & 18.24 $\pm$ 0.08 \\
7994 & Mephisto & $z$ & 17.39 $\pm$ 0.08 \\
7994 & Mephisto & $v$ & 20.28 $\pm$ 0.21 \\
8040 & ALT100C & $r$ & 18.37 $\pm$ 0.03 \\
8054 & Mephisto & $r$ & 18.37 $\pm$ 0.09 \\
8054 & Mephisto & $z$ & 17.55 $\pm$ 0.09 \\
8080 & TNOT & $gp$ & 19.13 $\pm$ 0.17 \\
8174 & Mephisto & $r$ & 18.29 $\pm$ 0.08 \\
8174 & Mephisto & $z$ & 17.47 $\pm$ 0.08 \\
8294 & Mephisto & $r$ & 18.37 $\pm$ 0.09 \\
8294 & Mephisto & $z$ & 17.65 $\pm$ 0.10 \\
8294 & Mephisto & $v$ & 19.71 $\pm$ 0.24 \\
8297 & ALT50D & $r$ & 18.33 $\pm$ 0.13 \\
8310 & TNOT & $rp$ & 18.37 $\pm$ 0.12 \\
8346 & ALT100C & $r$ & 18.38 $\pm$ 0.03 \\
8414 & Mephisto & $r$ & 18.33 $\pm$ 0.07 \\
8414 & Mephisto & $z$ & 17.71 $\pm$ 0.09 \\
8474 & Mephisto & $r$ & 18.36 $\pm$ 0.06 \\
8534 & Mephisto & $z$ & 17.68 $\pm$ 0.08 \\
8654 & Mephisto & $r$ & 18.51 $\pm$ 0.12 \\
8654 & Mephisto & $z$ & 17.49 $\pm$ 0.17 \\
8662 & ALT100C & $i$ & 18.08 $\pm$ 0.04 \\
8705 & ALT50D & $r$ & 18.44 $\pm$ 0.14 \\
8770 & TNOT & $gp$ & 19.10 $\pm$ 0.18 \\
8968 & ALT100C & $i$ & 18.04 $\pm$ 0.03 \\
9000 & TNOT & $rp$ & 18.62 $\pm$ 0.16 \\
9113 & ALT50D & $r$ & 18.43 $\pm$ 0.13 \\
9273 & ALT100C & $i$ & 18.11 $\pm$ 0.04 \\
9317 & ALT50D & $i$ & 18.24 $\pm$ 0.19 \\
9374 & Mephisto & $r$ & 18.58 $\pm$ 0.07 \\
9374 & Mephisto & $z$ & 17.81 $\pm$ 0.07 \\
9374 & Mephisto & $v$ & 21.09 $\pm$ 0.42 \\
9522 & ALT50D & $r$ & 18.55 $\pm$ 0.15 \\
9579 & ALT100C & $i$ & 18.21 $\pm$ 0.04 \\
9691 & TNOT & $rp$ & 18.44 $\pm$ 0.12 \\
9734 & Mephisto & $r$ & 18.55 $\pm$ 0.06 \\
9734 & Mephisto & $z$ & 17.97 $\pm$ 0.08 \\
9734 & Mephisto & $v$ & 20.35 $\pm$ 0.32 \\
9885 & ALT100C & $i$ & 18.16 $\pm$ 0.04 \\
9930 & ALT50D & $r$ & 18.51 $\pm$ 0.15 \\
10154 & Mephisto & $g$ & 18.96 $\pm$ 0.12 \\
10154 & Mephisto & $i$ & 18.11 $\pm$ 0.05 \\
10202 & ALT100C & $g$ & 19.54 $\pm$ 0.08 \\
10379 & TNOT & $rp$ & 18.93 $\pm$ 0.18 \\
10454 & Mephisto & $g$ & 18.90 $\pm$ 0.13 \\
10454 & Mephisto & $i$ & 18.06 $\pm$ 0.06 \\
10508 & ALT100C & $g$ & 19.45 $\pm$ 0.08 \\
10542 & ALT50D & $i$ & 18.12 $\pm$ 0.20 \\
10813 & ALT100C & $g$ & 19.57 $\pm$ 0.08 \\
10814 & Mephisto & $r$ & 18.77 $\pm$ 0.08 \\
10814 & Mephisto & $z$ & 17.92 $\pm$ 0.10 \\
10950 & ALT50D & $r$ & 18.89 $\pm$ 0.17 \\
11067 & TNOT & $rp$ & 18.77 $\pm$ 0.16 \\
11114 & Mephisto & $r$ & 18.58 $\pm$ 0.07 \\
11114 & Mephisto & $z$ & 18.05 $\pm$ 0.13 \\
11491 & ALT100C & $g$ & 19.57 $\pm$ 0.10 \\
11534 & Mephisto & $g$ & 19.29 $\pm$ 0.15 \\
11534 & Mephisto & $i$ & 18.28 $\pm$ 0.06 \\
11797 & ALT100C & $g$ & 19.83 $\pm$ 0.13 \\
11834 & Mephisto & $g$ & 19.33 $\pm$ 0.14 \\
11834 & Mephisto & $i$ & 18.27 $\pm$ 0.06 \\
12090 & ALT50D & $i$ & 17.90$^{*}$ \\
12194 & Mephisto & $r$ & 18.88 $\pm$ 0.08 \\
12194 & Mephisto & $z$ & 18.18 $\pm$ 0.14 \\
12419 & ALT50A & $z$ & 16.70$^{*}$ \\
12494 & Mephisto & $r$ & 19.16 $\pm$ 0.09 \\
12554 & Mephisto & $z$ & 17.88 $\pm$ 0.14 \\
12912 & ALT100C & $z$ & 18.24 $\pm$ 0.11 \\
14758 & ALT100C & $r$ & 19.50 $\pm$ 0.14 \\
15340 & ALT50A & $z$ & 16.90$^{*}$ \\
16602 & ALT100C & $i$ & 18.40$^{*}$ \\
28468 & NOT & $r$ & 20.41 $\pm$ 0.03 \\
29009 & NOT & $z$ & 19.73 $\pm$ 0.03 \\
29595 & NOT & $g$ & 21.34 $\pm$ 0.05 \\
30101 & NOT & $i$ & 20.02 $\pm$ 0.02 \\
\hline
\end{longtable} 
\end{center}
\end{ThreePartTable}

\begin{ThreePartTable}
\begin{center}
\begin{longtable}{cccc}
  \caption{Optical observations of GRB 250404A/EP250404a on clear filters. The magnitudes have not been corrected for foreground Galactic extinction. All errors represent the 1$\sigma$ uncertainties.}
  \label{tab:photoclear} \\
  \hline
  $T-T_0~(\rm s)$ & Telescope & Band & AB Magnitude \\
  \hline
  \endfirsthead 
  \multicolumn{4}{c}{{\textbf{\tablename\ \thetable{}.} Continued}} \\
  \hline
  $T-T_0~(\rm s)$ & Telescope & Band & AB Magnitude \\
  \hline
  \endhead
  \endlastfoot 
328 & HMT & Clear ($\sim$$G$) & 16.34 $\pm$ 0.06 \\
391 & HMT & Clear ($\sim$$G$) & 16.31 $\pm$ 0.04 \\
455 & HMT & Clear ($\sim$$G$) & 16.45 $\pm$ 0.04 \\
518 & HMT & Clear ($\sim$$G$) & 16.45 $\pm$ 0.04 \\
596 & HMT & Clear ($\sim$$G$) & 15.89 $\pm$ 0.02 \\
690 & HMT & Clear ($\sim$$G$) & 15.57 $\pm$ 0.01 \\
784 & HMT & Clear ($\sim$$G$) & 15.37 $\pm$ 0.01 \\
879 & HMT & Clear ($\sim$$G$) & 15.07 $\pm$ 0.01 \\
1003 & HMT & Clear ($\sim$$G$) & 14.90 $\pm$ 0.01 \\
1157 & HMT & Clear ($\sim$$G$) & 14.97 $\pm$ 0.01 \\
1203 & BOOTES & Clear ($\sim$$G$) & 14.82 $\pm$ 0.04 \\
1251 & Schmidt & Clear ($\sim$$G$) & 15.36 $\pm$ 0.01 \\
1310 & HMT & Clear ($\sim$$G$) & 15.17 $\pm$ 0.01 \\
1321 & Schmidt & Clear ($\sim$$G$) & 15.51 $\pm$ 0.01 \\
1390 & Schmidt & Clear ($\sim$$G$) & 15.59 $\pm$ 0.01 \\
1416 & BOOTES & Clear ($\sim$$G$) & 15.25 $\pm$ 0.04 \\
1460 & Schmidt & Clear ($\sim$$G$) & 15.82 $\pm$ 0.01 \\
1464 & HMT & Clear ($\sim$$G$) & 15.48 $\pm$ 0.01 \\
1529 & Schmidt & Clear ($\sim$$G$) & 15.92 $\pm$ 0.01 \\
1576 & BOOTES & Clear ($\sim$$G$) & 15.36 $\pm$ 0.07 \\
1599 & Schmidt & Clear ($\sim$$G$) & 15.86 $\pm$ 0.01 \\
1643 & HMT & Clear ($\sim$$G$) & 15.63 $\pm$ 0.01 \\
1669 & Schmidt & Clear ($\sim$$G$) & 15.97 $\pm$ 0.01 \\
1721 & BOOTES & Clear ($\sim$$G$) & 15.54 $\pm$ 0.05 \\
1736 & HMT & Clear ($\sim$$G$) & 15.78 $\pm$ 0.01 \\
1739 & Schmidt & Clear ($\sim$$G$) & 16.02 $\pm$ 0.01 \\
1809 & Schmidt & Clear ($\sim$$G$) & 16.08 $\pm$ 0.01 \\
1829 & HMT & Clear ($\sim$$G$) & 15.80 $\pm$ 0.01 \\
1878 & Schmidt & Clear ($\sim$$G$) & 16.06 $\pm$ 0.01 \\
1888 & BOOTES & Clear ($\sim$$G$) & 15.61 $\pm$ 0.05 \\
1922 & HMT & Clear ($\sim$$G$) & 15.93 $\pm$ 0.02 \\
1946 & Schmidt & Clear ($\sim$$G$) & 16.17 $\pm$ 0.01 \\
2016 & HMT & Clear ($\sim$$G$) & 15.96 $\pm$ 0.02 \\
2017 & Schmidt & Clear ($\sim$$G$) & 16.25 $\pm$ 0.02 \\
2063 & BOOTES & Clear ($\sim$$G$) & 15.75 $\pm$ 0.08 \\
2085 & Schmidt & Clear ($\sim$$G$) & 16.37 $\pm$ 0.02 \\
2109 & HMT & Clear ($\sim$$G$) & 16.12 $\pm$ 0.02 \\
2155 & Schmidt & Clear ($\sim$$G$) & 16.44 $\pm$ 0.02 \\
2204 & HMT & Clear ($\sim$$G$) & 16.18 $\pm$ 0.03 \\
2224 & Schmidt & Clear ($\sim$$G$) & 16.51 $\pm$ 0.02 \\
2293 & Schmidt & Clear ($\sim$$G$) & 16.48 $\pm$ 0.02 \\
2300 & HMT & Clear ($\sim$$G$) & 16.25 $\pm$ 0.02 \\
2353 & BOOTES & Clear ($\sim$$G$) & 16.17 $\pm$ 0.10 \\
2363 & Schmidt & Clear ($\sim$$G$) & 16.55 $\pm$ 0.02 \\
2394 & HMT & Clear ($\sim$$G$) & 16.33 $\pm$ 0.03 \\
2432 & Schmidt & Clear ($\sim$$G$) & 16.63 $\pm$ 0.02 \\
2487 & HMT & Clear ($\sim$$G$) & 16.41 $\pm$ 0.03 \\
2502 & Schmidt & Clear ($\sim$$G$) & 16.66 $\pm$ 0.02 \\
2571 & Schmidt & Clear ($\sim$$G$) & 16.78 $\pm$ 0.02 \\
2582 & HMT & Clear ($\sim$$G$) & 16.53 $\pm$ 0.03 \\
2640 & Schmidt & Clear ($\sim$$G$) & 16.89 $\pm$ 0.03 \\
2677 & HMT & Clear ($\sim$$G$) & 16.58 $\pm$ 0.03 \\
2708 & Schmidt & Clear ($\sim$$G$) & 16.95 $\pm$ 0.03 \\
2778 & Schmidt & Clear ($\sim$$G$) & 16.98 $\pm$ 0.03 \\
2800 & HMT & Clear ($\sim$$G$) & 16.68 $\pm$ 0.03 \\
2847 & Schmidt & Clear ($\sim$$G$) & 17.00 $\pm$ 0.03 \\
2916 & Schmidt & Clear ($\sim$$G$) & 17.03 $\pm$ 0.03 \\
2954 & HMT & Clear ($\sim$$G$) & 16.77 $\pm$ 0.03 \\
2985 & Schmidt & Clear ($\sim$$G$) & 17.12 $\pm$ 0.03 \\
3055 & Schmidt & Clear ($\sim$$G$) & 17.22 $\pm$ 0.04 \\
3067 & BOOTES & Clear ($\sim$$G$) & 16.64 $\pm$ 0.15 \\
3108 & HMT & Clear ($\sim$$G$) & 16.94 $\pm$ 0.04 \\
3125 & Schmidt & Clear ($\sim$$G$) & 17.18 $\pm$ 0.04 \\
3194 & Schmidt & Clear ($\sim$$G$) & 17.23 $\pm$ 0.04 \\
3262 & HMT & Clear ($\sim$$G$) & 16.96 $\pm$ 0.04 \\
3264 & Schmidt & Clear ($\sim$$G$) & 17.19 $\pm$ 0.03 \\
3333 & Schmidt & Clear ($\sim$$G$) & 17.21 $\pm$ 0.03 \\
3403 & Schmidt & Clear ($\sim$$G$) & 17.28 $\pm$ 0.04 \\
3472 & Schmidt & Clear ($\sim$$G$) & 17.35 $\pm$ 0.04 \\
3517 & HMT & Clear ($\sim$$G$) & 17.00 $\pm$ 0.04 \\
3540 & Schmidt & Clear ($\sim$$G$) & 17.35 $\pm$ 0.04 \\
3610 & Schmidt & Clear ($\sim$$G$) & 17.42 $\pm$ 0.05 \\
3670 & HMT & Clear ($\sim$$G$) & 17.15 $\pm$ 0.04 \\
3679 & Schmidt & Clear ($\sim$$G$) & 17.63 $\pm$ 0.05 \\
3748 & Schmidt & Clear ($\sim$$G$) & 17.38 $\pm$ 0.04 \\
3818 & Schmidt & Clear ($\sim$$G$) & 17.54 $\pm$ 0.05 \\
3823 & HMT & Clear ($\sim$$G$) & 17.31 $\pm$ 0.04 \\
3840 & BOOTES & Clear ($\sim$$G$) & 17.21 $\pm$ 0.19 \\
3887 & Schmidt & Clear ($\sim$$G$) & 17.57 $\pm$ 0.05 \\
3957 & Schmidt & Clear ($\sim$$G$) & 17.62 $\pm$ 0.05 \\
3977 & HMT & Clear ($\sim$$G$) & 17.38 $\pm$ 0.04 \\
4026 & Schmidt & Clear ($\sim$$G$) & 17.66 $\pm$ 0.05 \\
4095 & Schmidt & Clear ($\sim$$G$) & 17.73 $\pm$ 0.06 \\
4130 & HMT & Clear ($\sim$$G$) & 17.49 $\pm$ 0.04 \\
4165 & Schmidt & Clear ($\sim$$G$) & 17.74 $\pm$ 0.06 \\
4233 & Schmidt & Clear ($\sim$$G$) & 17.88 $\pm$ 0.07 \\
4284 & HMT & Clear ($\sim$$G$) & 17.51 $\pm$ 0.05 \\
4304 & Schmidt & Clear ($\sim$$G$) & 18.01 $\pm$ 0.08 \\
4372 & Schmidt & Clear ($\sim$$G$) & 17.77 $\pm$ 0.06 \\
4438 & HMT & Clear ($\sim$$G$) & 17.46 $\pm$ 0.04 \\
4442 & Schmidt & Clear ($\sim$$G$) & 17.83 $\pm$ 0.07 \\
4512 & Schmidt & Clear ($\sim$$G$) & 17.93 $\pm$ 0.07 \\
4581 & Schmidt & Clear ($\sim$$G$) & 17.84 $\pm$ 0.07 \\
4651 & Schmidt & Clear ($\sim$$G$) & 17.88 $\pm$ 0.07 \\
4722 & Schmidt & Clear ($\sim$$G$) & 18.00 $\pm$ 0.07 \\
4791 & Schmidt & Clear ($\sim$$G$) & 18.04 $\pm$ 0.08 \\
4862 & Schmidt & Clear ($\sim$$G$) & 17.94 $\pm$ 0.07 \\
4933 & Schmidt & Clear ($\sim$$G$) & 18.15 $\pm$ 0.09 \\
4948 & HMT & Clear ($\sim$$G$) & 17.72 $\pm$ 0.05 \\
5003 & Schmidt & Clear ($\sim$$G$) & 18.07 $\pm$ 0.09 \\
5073 & Schmidt & Clear ($\sim$$G$) & 18.13 $\pm$ 0.09 \\
5143 & Schmidt & Clear ($\sim$$G$) & 18.11 $\pm$ 0.09 \\
5161 & HMT & Clear ($\sim$$G$) & 17.79 $\pm$ 0.05 \\
5588 & HMT & Clear ($\sim$$G$) & 17.95 $\pm$ 0.07 \\
5801 & HMT & Clear ($\sim$$G$) & 18.03 $\pm$ 0.08 \\
6015 & HMT & Clear ($\sim$$G$) & 17.98 $\pm$ 0.07 \\
6379 & Schmidt & Clear ($\sim$$G$) & 18.12 $\pm$ 0.04 \\
6442 & HMT & Clear ($\sim$$G$) & 18.42 $\pm$ 0.13 \\
6655 & HMT & Clear ($\sim$$G$) & 18.16 $\pm$ 0.09 \\
6869 & HMT & Clear ($\sim$$G$) & 18.14 $\pm$ 0.10 \\
7327 & HMT & Clear ($\sim$$G$) & 18.31 $\pm$ 0.10 \\
7561 & HMT & Clear ($\sim$$G$) & 18.58 $\pm$ 0.12 \\
7863 & HMT & Clear ($\sim$$G$) & 18.57 $\pm$ 0.11 \\
8291 & HMT & Clear ($\sim$$G$) & 18.54 $\pm$ 0.12 \\
8505 & HMT & Clear ($\sim$$G$) & 18.79 $\pm$ 0.14 \\
8719 & HMT & Clear ($\sim$$G$) & 18.43 $\pm$ 0.11 \\
8932 & HMT & Clear ($\sim$$G$) & 18.62 $\pm$ 0.15 \\
9146 & HMT & Clear ($\sim$$G$) & 18.70 $\pm$ 0.13 \\
9573 & HMT & Clear ($\sim$$G$) & 18.80 $\pm$ 0.11 \\
\hline
\end{longtable} 
\end{center}
\end{ThreePartTable}

\section{Afterglow Fitting Corner Plot}

\begin{figure*}
 \centering
 \includegraphics[width = 0.9\textwidth]{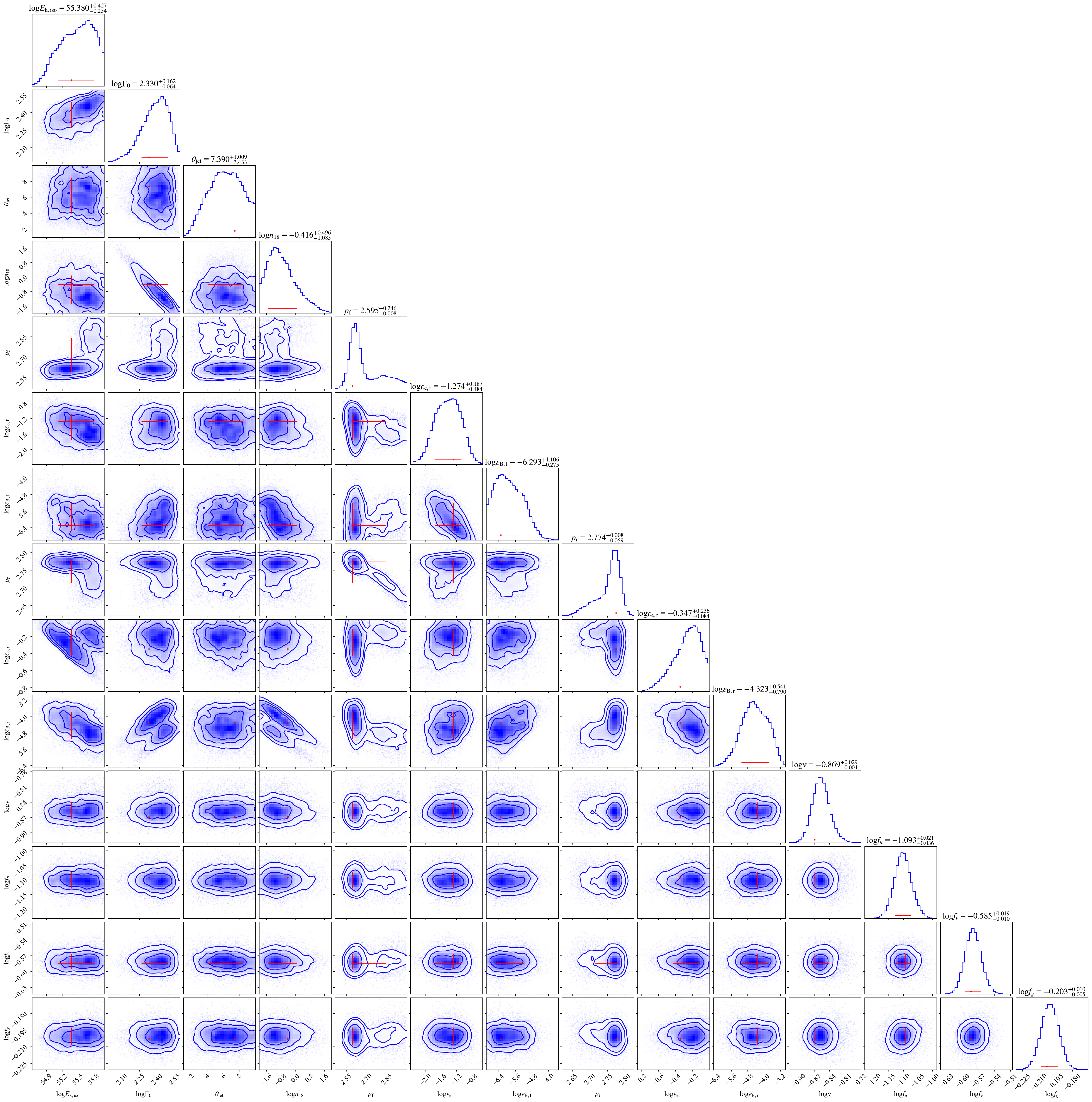}
 \caption{Corner plot of the posterior probability distributions of the parameters for afterglow fitting using the FS+RS model with $u$-, $v$-, and $g$-band correction factors. The red error bars represent 1$\sigma$ uncertainties.}
 \label{fig:corner}
\end{figure*}

Appendix C contains the corner plot in Figure \ref{fig:corner} showing the posterior probability distributions of the parameters from the FS+RS afterglow fitting with $u$-, $v$-, and $g$-band correction factors.
 
\end{document}